%% file: DFTpOFDMA_red_v2.tex
\documentclass[journal, onecolumn]{IEEEtran}

\usepackage{amsmath,amssymb}
\usepackage{graphicx}
\usepackage{epsfig}
\usepackage{tikz}
\usetikzlibrary{shapes,shadows,arrows}

\begin{document}

\title{Limiting Performance of Conventional and Widely Linear DFT-precoded-OFDM Receivers in Wideband Frequency Selective Channels}
\author{
\authorblockN{Kiran Kuchi}\\
\authorblockA{Indian Institute of Technology Hyderabad, India} \\
}
\maketitle
\begin{abstract}
This paper describes the limiting behavior of linear and decision feedback equalizers (DFEs) in single/multiple antenna systems employing real/complex-valued modulation alphabets. The wideband frequency selective channel is modeled using a Rayleigh fading channel model with infinite number of time domain channel taps. Using this model, we show that the considered equalizers offer a fixed post signal-to-noise-ratio (post-SNR) at the equalizer output that is close to the matched filter bound (MFB). General expressions for the post-SNR are obtained for zero-forcing (ZF) based conventional receivers as well as for the case of receivers employing widely linear (WL) processing. Simulation is used to study the bit error rate (BER) performance of both MMSE and ZF based receivers. Results show that the considered receivers advantageously exploit the rich frequency selective channel to mitigate both fading and inter-symbol-interference (ISI) while offering a performance comparable to the MFB.
\end{abstract}

\input formulaDef_08Aug2007.tex

\section{Introduction}
Linear and decision feedback equalizers (DFEs) have been widely studied for the past 50 years. In particular, these two receiver structures have been extensively used for application in narrow-band cellular radio systems such as IS-54 and GSM. With the introduction of Discrete Fourier transform-precoded-OFDMA (DFT-precoded-OFDMA) \cite{LTEul},\cite{ETT:scfdma1} in the uplink of the long-term-evolution (LTE) standard \cite{3p9G:standard}, there has been renewed interest in the design and analysis of these two receivers operating in wideband frequency selective channels. DFT-precoded-OFDM, also known as single carrier FDMA (SC-FDMA), is a variant of OFDM in which the modulation data is precoded using the DFT before transmitting the data on the subcarriers.  The resultant modulation signal exhibits single carrier properties such as sinc \cite{Proakis:proakis} like pulse shaping and low peak-to-average-power-ratio (PAPR). As the DFT precoder introduces inter-symbol interference (ISI), this method requires sophisticated channel equalization at the receiver. \\

In broadband wireless systems employing high bandwidths, the propagation channel typically exhibits high frequency selectivity. For these systems, link performance measures such as the diversity order and bit-error-rate (BER) of a conventional Minimum mean-square estimation (MMSE) based linear equalziers have not yet been fully characterized \cite{LE1:Aug08}\textendash\nocite{LE2:Jan04}\nocite{LE1:Aug08}\nocite{LE3:Nov04}\nocite{Slock:2004}\nocite{Lataief:2004}\cite{Giannakis:2004}. The noise enhancement phenomenon which is inherent in linear equalizers poses a difficulty in analyzing the receiver performance. The MMSE-DFE \cite{Cioffi:Nov1}, \cite{Cioffi:Nov2}, on the other hand, is an optimum canonical receiver for channels with ISI. In frequency selective channels, it provides full diversity, and performance is generally comparable to the optimum matched filter bound (MFB). However, the performance gap between MMSE-DFE and the optimum MFB has not yet been analytically determined for frequency selective fading channels. In many cases, simulation is typically used to determine the link performance. \\

The performance loss caused by the decision feedback section of the MMSE-DFE can be minimized by using a receiver structure that uses the MMSE-DFE feed-forward filter as a pre-filter \cite{AlDhahir:July95} which provides a minimum phase response followed by a reduced state sequence estimation (RSSE) \cite{Eyuboglu:Jan88} algorithm that uses set partitioning and state dependent decision feedback principles; the maximum likelihood sequence estimator (MLSE) \cite{Forney:May72},\cite{Ungerboeck:74} can be viewed as a special case of RSSE. In typical channels, RSSE with an appropriately chosen number of states performs close to MLSE \cite{Gerstacker:Jan02}. In spite of the availability of a number of alternatives to MLSE, linear and decision feedback equalizers are generally preferred in wide-band systems due to low implementation complexity \cite{3p9G:standard}.\\

In DFT-precoded-OFDM systems, the MMSE-DFE \cite{Cioffi:Stanford}\textendash\nocite{AlDhahir:July95}\nocite{Winters:July99}\cite{Fischer} equalizer can be implemented efficiently using a frequency domain feed-forward filter (FFF) followed by a time domain DFE \cite{Italia:02}\textendash\nocite{Benvenuto:05}\nocite{Benvenuto:05}\nocite{Padmanabhan:Jan09}\nocite{Prasad:09}\nocite{Lin}\cite{Gerstacker:may08}. Computation of FFF and feedback filters (FBF) for DFT-precoded-OFDM differs from conventional  single carrier methods. The circular symmetry inherent to OFDM simplifies the computational requirements of both filter calculation and implementation. In \cite{Italia:02}, \cite{Falconer:02} simulation is used in to show that the coded bit error rate (BER) of DFT-precoded-OFDM is comparable to standard OFDM for single antenna systems.\\

For real-valued data transmission (e.g., binary-phase-shift-keying (BPSK) or Amplitude-shift-keying (ASK)), widely linear (WL) equalizers which jointly filter the received signal and its complex-conjugate \cite{picinbono:95} are known to outperform conventional receivers. This concept has been applied for numerous wireless applications \cite{YoonLeib:Jan97}\textendash~\nocite{Gelli:Jun00}\nocite{Buzzi:June00}\nocite{Lampe:Sept01}\nocite{Lampe:Feb02}\nocite{Gerstacker:Jan02}\nocite{Gerstacker:Sept03}\nocite{Gerstacker:July04}\nocite{Gelli:Mar05}\nocite{Gelli:Mar05}\cite{Chevaliar:Mar2006}
including equalization, interference suppression, multi-user detection etc. Implementation WL equalizers is discussed  in \cite{Gerstacker:Sept03} for conventional time domain single carrier systems where simulation is used to show that WL processing reduces the noise enhancement as well as an increase in SNR at the equalizer output compared to conventional equalizers. WL receiver algorithms are widely employed in GSM for a) low-complexity equalization of binary GMSK modulation in frequency selective channels b) co-channel interference suppression using a single receiver antenna. This latter feature is popularly known as single antenna interference cancelation (SAIC) \cite{Chevaliar:Mar2006}-\cite{Trigui:may01}.\\

Throughput this paper, we assume that the receiver has multiple spatially separated antennas. However, the analysis and the results of this paper hold for the case of single antenna as well. Typical wideband frequency selective channel models such as Ped-B and Veh-A channels \cite{EVM} have a large number of multipath fading taps with independent Rayleigh fading. Performance of the receiver in these types of channels can be well approximated by using a wideband channel with infinite amount of channel memory \cite{Kuchi:2012}. In particular, we consider a channel with $v$ time domain taps where the individual taps are modeled as independent and identically distributed (i.i.d.) complex Gaussian random variables with zero-mean with per tap variance of $\frac{1}{v}$. The post-processing signal-to-noise-power-ratio (post-SNR) of the considered equalizers is analyzed in the limiting case as $v \rightarrow \infty$. Using this model, \cite{Kuchi:2012} has shown that the SNR at the output of a multi-antenna ZF-LE with $N_r$ antennas reaches a constant mean value of $\frac{N_r-1}{\sigma^2_{n}}$ where $\sigma^2_{n}$ denotes the noise variance and $N_r>1$. For the case of single receiver antenna, both ZF-LE and MMSE-LE are shown to perform poorly. Therefore, it is worthwhile to consider the DFE as an implementation alternative. \\

In this paper, we further generalize the results of  \cite{Kuchi:2012} and analyze the limiting performance of three receiver algorithms, namely, a) conventional ZF-DFE b) WL ZF-LE c) WL ZF-DFE. While ZF based methods permit analytical evaluation of the post-SNR of the receiver, simulation is used to study the performance of MMSE based receivers. The post-SNR bounds developed in this paper provides new insights into the receiver performance. Specifically, we show that in i.i.d fading channels with infinitely high frequency selectivity, the post-SNR at the output of all the considered receivers reach a fixed SNR. Using these results, we quantify the performance gap of a given receiver with respect to the MFB. \\

The organization of the paper is as follows. In section \ref{conmmsedfe}, we first generalize the finite length ZF/MMSE-DFE results to the infinite length case. Then, we obtain a general expression for the post-SNR of a ZF-DFE for the case of infinite length i.i.d. fading channel under the assumption of error free decision feedback (ideal DFE). In section \ref{WL equalizer}, we present the limiting analysis for receivers employing WL processing. Collection of complex and complex-conjugated copies of the received signal effectively doubles the number of receiver branches. We show that these additional degrees-of-freedom (DOF)\footnote{Here DOF denotes the total number of receiver branches} obtained through WL processing helps the receiver to obtain a substantially higher post-SNR compared to conventional LEs. Analogous to the case of conventional ZF-DFE, in section \ref{wlmmsedfe}, we obtain filter settings for WL ZF/MMSE-DFE receiver. Then a general expression for the post-SNR of the WL ZF-DFE is obtained for the case of infinite length i.i.d. fading channel. In section \ref{results}, we  present simulation results which illustrate the DFE error propagation effects. Finally, conclusions are drawn in section \ref{conclusions}.\\

\texttt{Notation}\\
The following notation is adopted throughout the paper. Vectors are denoted using bold-face lower-case letters, matrices are denoted using bold-face upper-case letters. Time domain quantities are denoted using the subscript $t$. The $M$-point DFT of a vector $\bh_t(l)$ is defined as: $\bh(k)= \sum_{l=0}^{M-1} \bh_t(l)e^{\frac{-j 2 \pi kl}{M}}$ where $k=0,1,..,M-1$. The corresponding $M$-point IDFT is given by: $\bh_t(l)= \frac{1}{M}\sum_{k=0}^{M-1} \bh(k)e^{\frac{-j 2 \pi kl}{M}}$. The squared Euclidean norm of a row/column vector $\bh(k)=[h_1(k), h_2(k),..,h_n(k)]$ is denoted as: $||\bh(k)||^2=\sum_{m=1}^{n} |h_m(k)|^2$. The circular convolution between two length $N$ sequences is defined as: $x_1(n) \odot x_2(n)= \sum_{n=0}^{N-1} x_1(n) x_2((m-n))_{N}$ where the subscript in $x_2((m-n))_{N}$ denotes modulo $N$ operation and  $\odot$ denotes circular convolution operation. The symbols $\dagger$, $*$, $Tr$ denote Hermitian, complex-conjugate and transpose operations, respectively and $E[.]$ denotes expectation operator.
\section{System Model}\label{sysmodel}
The DFT-precoded-OFDMA transmitter sends a block of $M$ i.i.d. real/complex-valued modulation alphabets with zero-mean and  variance of $\sigma^2_x$. The DFT precoding of the data stream $x_t(l)$ is accomplished using a $M$-point Fast Fourier Transform (FFT) as
\begin{equation}\label{Eqn1}
    x(k) =  \sum_{l=0}^{M-1} x_t(l) e^{\frac{-j2 \pi l k}{M} }, \quad k=0,..,M-1
\end{equation}
where $l,k$  denote the discrete time and subcarrier indices, respectively. Throughout this paper we consider wideband allocation. Therefore, the precoded data is mapped to all the available $M$ contiguous subcarriers. The time domain baseband signal $s(t)$ is obtained using an inverse discrete time Fourier transform (IDTFT)
\begin{equation}\label{Eqn2}
    s(t)= \frac{1}{M} \sum_{k=0}^{M-1} x(k) e^{j2 \pi k \Delta f (t-T_{CP})}, \quad t\in [0, T+T_{CP}]
\end{equation}
where $T$ is the useful portion of OFDMA symbol, $T_{CP}$ is the duration of the cyclic prefix (CP) and $\Delta f=\frac{1}{T}$ is the subcarrier spacing.
\section{MMSE-DFE Receiver}\label{conmmsedfe}
The receiver front end operations such as sampling, synchronization, CP removal and channel estimation operations are similar to a conventional system. Further, the memory introduced by the propagation channel is assumed to be less than that of the CP duration. Throughout this paper, ideal knowledge of channel state information is assumed at the receiver. We consider a receiver equipped with $N_r$ antennas. Stacking up the time domain sample outputs of multiple receiver antennas in a column vector format, we get
\begin{equation}\label{Eqn2aa}
    \by_t(l)= s_t(l) \odot \bh_t(l)+\bn_t(l), \quad l=0,1,..,M-1
\end{equation}
where
\begin{eqnarray*}
    \by_t(l)&=& [y_{t,1}(l),..,y_{t,N_r}(l)]^{Tr}, \, \,\bh_t(l)= [h_{t,1}(l),..,h_{t,N_r}(l)]^{Tr},\\
    \bn_t(l)&=&[n_{t,1}(l),..,n_{t,N_r}(l)]^{Tr}
\end{eqnarray*}
denote the received signal, channel, and noise vectors of size $N_r \times 1$. Here, $s_t(l)$ corresponds to the sampled version of the analog signal $s(t)$. The noise vector $\bn_t(l)$ is composed of $N_r$ i.i.d. complex-Gaussian noise random variables each with zero-mean and variance $\frac{\sigma^{2}_n}{2}$ per dimension. Note that $\bh_t(n)$ is assumed to be a time limited channel vector where each element of $\bh_t(n)$ has a duration $v$ samples and $M >> v$. Taking the $M$-point DFT of $ \by_t(l)$, we get
\begin{eqnarray}
\by(k) &=& \bh(k) x(k) +\bn(k), \quad k=0,1,..,M-1\label{Eqna}
\end{eqnarray}
where $\by(k)=DFT[\by_t(l)]$, $\bh(k)=DFT[\bh_t(l)]$, $x(k)=DFT[s_t(l)]$, $\bn(k)=DFT[\bn_t(l)]$. In the MMSE-DFE receiver (see Fig \ref{fig:convmmsedfe}), the received signal is filtered using a vector-valued feed-forward filter to obtain: $z(k)= \bw(k) \by(k)$. Let $\vec{z}(k)=z(k)-b(k)x(k)$ is the ISI free signal where $b(k)$ is the frequency domain FBF. Here, $1+b(k)=1+\sum_{l=1}^{L} b_t(l) e^{\frac{-j 2 \pi kl}{M}}$ where the FBF is constrained to have $L$ time domain taps. Note that $L$ is a receiver design parameter and its value can be chosen to be equal to the channel memory. As shown in Fig \ref{fig:convmmsedfe}, the FBF is implemented in time domain to obtain a decision variable
\begin{eqnarray}
\vec{z}_t(l)= z_t(l)-\underbrace{\sum_{m=1}^{L} b_t(m) x_t(l \ominus m)}_{\texttt{ISI}}. \label{errort1}
\end{eqnarray}
where $z_t(l)$ is obtained after taking the IDFT of $z(k)$ and $\vec{z}_t(l)$ is the ISI free time domain signal which is fed to the symbol demodulator. Here, the symbol $\ominus$ denotes left circular shift operation.\\
MMSE-DFE filter expressions are given in \cite{Italia:02} for single receiver antenna case and results for MIMO systems are available in \cite{Gerstacker:may08}. In order to obtain an expression for the mean-square error (MSE) which enables closed-form analysis in i.i.d. fading channels, in Appendix-1 we first provide expressions for the MMSE-DFE filter for the finite length case, then we generalize the results for the infinite length scenario. Using these results, for $M \rightarrow \infty$, the post-SNR defined as the SNR at the output of an un-biased MMSE-DFE receiver is given by
\begin{eqnarray}
\texttt{SNR}_{\texttt{MMSE-DFE}, M \rightarrow \infty} &=& e^{\lim_{M \rightarrow \infty}\frac{1}{M}\sum_{k=0}^{M-1}\ln \left[\frac{\sigma^{2}_x||\bh(k)||^2+\sigma^{2}_n}{\sigma^{2}_n}\right]}-1. \label{post-snrf1}
\end{eqnarray}
\subsection{Limiting performance of ZF-DFE in wideband channels}\label{ZFDFEiid}
Recognizing that the ZF-DFE enables one to obtain a closed-form expression for the post-SNR, we analyze its behavior for the proposed i.i.d. fading channel with infinite length. To this end, let
\begin{eqnarray}
\bh(k)=\sum_{l=0}^{v-1} \bh_t(l) e^{\frac{-j2 \pi k l}{M}}, \quad k=0,1,..,M-1 \label{DFTchannel}
\end{eqnarray}
where $v$ is the effective channel length.  We are interested in determining the performance of the link for the limiting case where $v\rightarrow \infty$. Note that as $v$ tends to $\infty$, since $M >> v$, $M$ also tends to $\infty$. Therefore,
\begin{eqnarray}
\bh(k) &=& \lim_{v \rightarrow \infty, M \rightarrow \infty}\sum_{l=0}^{v-1} \bh_t(l) e^{\frac{-j2 \pi k l}{M}}\\
       &=& \lim_{M \rightarrow \infty}\sum_{l=0}^{M} \bh_t(l) e^{\frac{-j2 \pi k l}{M}}.  \label{DFTchannel_1}
\end{eqnarray}
Note that the variable $v$ is replaced with $M$ in (\ref{DFTchannel_1}) line 2 because as $v \rightarrow \infty$, $M \rightarrow \infty$, since $v << M$. Next, we model $\bh_t(l)$ as an i.i.d. zero-mean, complex-Gaussian vector with covariance $E\left(\bh_t(l)\bh^{\dagger}_t(l)\right)=\frac{\bI}{v}$. Note that per-tap power is set to $\frac{1}{v}$ so that the total power contained in the multi-path channel becomes unity. As $v \rightarrow \infty$, we can express the covariance term as: $\lim_{v \rightarrow \infty} E\left(\bh_t(l)\bh^{\dagger}_t(l)\right)=\lim_{v \rightarrow \infty}\frac{\bI}{v}=\lim_{M \rightarrow \infty}\frac{\bI}{M}$. Again here, $v$ is replaced with $M$ in the limit as $v \rightarrow \infty$. We have an infinite number of taps with vanishingly small power. However, the sum total power of all the taps is equal to unity. Using (\ref{DFTchannel_1}), it can be shown that $\bh(k)$ approaches an  i.i.d. complex-Gaussian vector with zero-mean and the covariance tends to an identity matrix i.e.,  $\lim_{v \rightarrow \infty}E\left(\bh(k)\bh^{\dagger}(k)\right) \rightarrow \bI$. More specifically, the probability density function of the elements of the channel vector $\bh(k)$ approaches an i.i.d. complex-Gaussian distribution with zero-mean, unit variance and the vectors $\bh(k)$ become statistically independent for $k=0,1,..,M-1$.\\
By setting $\sigma^{2}_n=0$ in the numerator of (\ref{post-snrf1}), we obtain the post-SNR of a ZF-DFE as
\begin{eqnarray}
\texttt{SNR}_{\texttt{ZF-DFE}}  &=& \frac{\sigma^{2}_x}{\sigma^{2}_n}e^{\lim_{M \rightarrow \infty}\frac{1}{M}\sum_{k=0}^{M-1}\ln||\bh(k)||^2}.
\end{eqnarray}
Applying central limit theorem (CLT), the r.v., $\lim_{M \rightarrow \infty}\frac{1}{M}\sum_{k=0}^{M-1}\ln ||\bh(k)||^2$ approaches Gaussian distribution with mean
\begin{eqnarray}
E \left[\lim_{M \rightarrow \infty}\frac{1}{M}\sum_{k=0}^{M-1}\ln ||\bh(k)||^2 \right] =E \left[\ln ||\bh(k)||^2 \right]
\end{eqnarray}
and variance
\begin{eqnarray}
\texttt{Var} \left[\lim_{M \rightarrow \infty}\frac{1}{M}\sum_{k=0}^{M-1}\ln ||\bh(k)||^2\right] = \lim_{M \rightarrow \infty} \frac{1}{M}\texttt{Var} \left[\ln ||\bh(k)||^2\right].
\end{eqnarray}
The expected logarithm of a chi-square random variable with $2N_r$ DOF is \cite{Paul:03}: $E \left[\ln ||\bh(k)||^2\right]=\left(-\beta+\sum_{m=1}^{N_r-1} \frac{1}{m}\right)$ where $\beta=0.577$ is the Euler's constant, and the variance is \cite{Paul:03}: $\texttt{Var}\left[ \ln ||\bh(k)||^2\right]=\sum_{p=1}^{\infty}\frac{1}{(p+N_r-1)}$. Since the variance term takes a finite value (the series is absolutely convergent), the variance of $\lim_{M \rightarrow \infty}\frac{1}{M}\sum_{k=0}^{M-1}\ln ||\bh(k)||^2$ approaches zero. Therefore, the SNR at the output of the ZF-DFE approaches a constant value of
\begin{eqnarray}
\texttt{SNR}_{\texttt{ZF-DFE}} &=&  \frac{\sigma^{2}_x e^{\left(-\beta+\sum_{m=1}^{N_r-1} \frac{1}{m}\right)}}{\sigma^{2}_n} \label{ZF DFE snr}
\end{eqnarray}
Note that the ZF-LE provides a fixed mean SNR of \cite{Kuchi:2012}
\begin{eqnarray}
\texttt{SNR}_{\texttt{ZF-LE}} &=&  \frac{\sigma^{2}_x (N_r-1)} {\sigma^{2}_n},  \quad \texttt{for} \quad N_r>1 \label{ZF-LE snr}
\end{eqnarray}
whereas the post-SNR corresponding to the MFB is given by: $\texttt{SNR}_{\texttt{MFB}}=\frac{N_r}{\sigma^2_{n}}$. \\
The above result suggests that highly dispersive nature of the frequency selective channels can be exploited advantageously to obtain a performance comparable to the MFB. After evaluating the expression (\ref{ZF DFE snr}) for the case of a single receiver antenna case, the ZF-DFE provides a post-SNR of $\frac{0.5616}{\sigma^2_{n}}$ which is 2.5 dB less than the MFB. For this case, both ZF and MMSE based LEs perform poorly compared to the MFB \cite{Kuchi:2012}. However, the ZF-DFE does not suffer from this limitation and provides a substantial gain over MMSE/ZF-LE. For $N_r=2$, the loss of ZF-DFE with respect to the MFB reduces to 1.19 dB whereas the ZF-LE has a higher loss of 3.0 dB.
\subsection{DFE initialization} \label{dfeinit}
In the MMSE-DFE implementation considered in this paper, the feedback filter is implemented in time domain. In (\ref{errort1}), the ISI term $\sum_{m=1}^{L} b_t(m) x_t(l \ominus m)$ is obtained by circularly convolving FBF $b_t(l)$ with the data sequence $x_t(l)$. For detecting the first data symbol $x_t(0)$, the receiver has to eliminate the ISI caused by the last $L$ data symbols of the data sequence $x_t(l)$. Specifically, the DFE requires knowledge of the data symbols $\bx_i=[x_t(N-L),..,x_t(N-2),x_t(N-1)]$. As proposed in \nocite{Padmanabhan:Jan09}, we use a linear equalizer to obtain hard decisions for required elements contained in $\bx_i$. These symbol estimates are then used to initialize the DFE. Simulation shows that this approach works quite well and the loss in the performance compared to the case of an ideal DFE is acceptable. We would like to remark here that an iterative receiver is presented in \cite{Benvenuto:05} to address the DFE initialization problem. The results of this paper show that MMSE-LE based initialization is sufficient to obtain near ideal performance. An alternative receiver initialization method is also discussed in \cite{Gerstacker:may08} for trellis based receivers.
\section{Widely Linear Frequency Domain MMSE Equalizer}\label{WL equalizer}
For the special case of real-constellations, we consider a frequency domain widely linear equalizer which jointly filters the complex-valued received signal and its complex-conjugated and frequency reversed copy in frequency domain (see Fig \ref{fig:wlmmsele}). Recall that the frequency domain signal model is given by (\ref{Eqna})
\begin{eqnarray}
\by(k) &=& \bh(k) x(k) +\bn(k), \quad k=0,1,..,M-1.
\end{eqnarray}
Applying complex conjugation and frequency reversal operation on $\by(k)$ we get
\begin{eqnarray}
\by^{*}(M-k) &=& \bh^{*}(M-k)x^{*}(M-k) + \bn^{*}(M-k) \\ \label{Eqnb}
&=& \bh^{*}(M-k)x(k) + \bn^{*}(M-k),  \quad k=0,1,..,M-1
\end{eqnarray}
where we use the fact that $x^{*}(M-k)=x(k)$ for real-valued modulation data. Combining (\ref{Eqna}), (\ref{Eqnb}) in vector form, we have
\begin{eqnarray}
   \left[ \begin{array}{c}
  \by(k) \\
  \by^{*}(M-k)\\
\end{array}\right] =  \left[ \begin{array}{c}
  \bh(k) \\
   \bh^{*}(M-k)\\
\end{array}\right]x(k)+\left[ \begin{array}{c}
 \bn(k) \\
 \bn^{*}(M-k)\\
\end{array}\right]
\end{eqnarray}.
We note that two copies of the frequency domain modulation signal $x(k)$ are obtained with distinct channel coefficients. Using compact vector notation,
$
  \bar{\by}(k)=\bar{\bh}(k)x(k)+ \bar{\bn}(k).
$
The WL filter $\bar{\bw}(k)=\left[\bw(k), \bw^{*}(M-k)\right]$ jointly filters the frequency domain signal $\by(k)$, and its complex-conjugated and frequency reversed copy $\by^{*}(M-k)$ to obtain the scalar decision variable denoted as $\bar{z}(k)$. Let $ \bar{z}(k)=\bw(k)\by(k)+\bw^{*}(M-k)\by^{*}(M-k)=\bar{\bw}(k) \bar{\by}(k)$.  An estimate of the desired data is obtained as: $\bar{z}_t(l)=IDFT[\bar{z}(k)]$. Using standard MMSE estimation \cite{Cioffi:Stanford}, the vector-valued WL MMSE filter is given by
\begin{eqnarray}
  \bar{\bw}(k)&=&\frac{\bar{\bh}^{\dagger}(k)}{1+\frac{\sigma^{2}_n}{\sigma^{2}_x} \bar{\bh}^{\dagger}(k)\bar{\bh}(k)}
\end{eqnarray}
Since $\bar{\bw}(k)=[\bw(k), \bw^{*}(M-k)]$ where $\bw(k)=\frac{\bh^{*}(k)}{1+\frac{\sigma^{2}_n}{\sigma^{2}_x} \bar{\bh}^{\dagger}(k)\bar{\bh}(k)}$, it is computationally efficient to calculate the filter $\bw(k)$ explicitly. The filter $\bw^{*}(M-k)$ can be obtained from $\bw(k)$ with low computational complexity using complex-conjugation and frequency reversal operations. The minimum MSE for this case is expressed as
\begin{eqnarray}
 \texttt{ MSE}_{\texttt{WL MMSE}}= \frac{1}{M} \sum_{k=0}^{M-1}  \frac{\sigma^{2}_n \sigma^{2}_x}{\sigma^{2}_n+\sigma^{2}_x \bar{\bh}^{\dagger}(k)\bar{\bh}(k)}. \label{Eq:6}
\end{eqnarray}
Note that
\begin{eqnarray*}
\bar{\bh}^{\dagger}(k)\bar{\bh}(k) &=& ||\bh(k)||^2+||\bh(M-k)||^2.
 \end{eqnarray*}
Using this result, the MSE can be expressed as
\begin{eqnarray}
\texttt{MSE}_{\texttt{WL MMSE}} &=& \frac{1}{M} \sum_{k=0}^{M-1} \left[ \frac{\sigma^{2}_n }{\frac{\sigma^{2}_n}{\sigma^{2}_x}+\left(||\bh(k)||^2+||\bh(M-k)||^2\right)} \right]. \label{EqnMSE}
\end{eqnarray}
The post-SNR defined as the SNR at the output of the WL MMSE receiver is given by
\begin{eqnarray}
\texttt{SNR}_{\texttt{WL MMSE}} = \frac{\sigma^{2}_x}{\texttt{MSE}_{\texttt{WL MMSE}} } = \frac{\sigma^{2}_x}{\sigma^{2}_n D}
\end{eqnarray}
where $D=\frac{1}{M}  \sum_{k=0}^{M-1} \left[\frac{1}{\frac{\sigma^{2}_n}{\sigma^{2}_x}+\left(||\bh(k)||^2+||\bh(M-k)||^2\right)}\right] $.
\subsection{Liming performance of WL ZF-LE}\label{wlzfleiid}
To obtain a closed-form expression for the post-SNR at the output of the equalizer, we analyze the performance of a ZF WL-LE. Letting $\sigma^{2}_n=0$ in the denominator of $D$, for $v \rightarrow \infty$, we let $M \rightarrow \infty$. In this case, we get
\begin{eqnarray}
D &=&  \lim_{M \rightarrow \infty} \frac{1}{M} \sum_{k=0}^{M-1} \left[\frac{1}{ \left(||\bh(k)||^2+||\bh(M-k)||^2\right)}\right]. \label{Deno}
\end{eqnarray}
Let $Q(k)=||\bh(k)||^2+||\bh(M-k)||^2$. Then we have: $Q(k)=Q(M-k)$, $Q(0)=2||\bh(0)||^2$, and $Q(\frac{M}{2})=2||\bh(\frac{M}{2})||^2$. Using this,
\begin{eqnarray}
D &=& \lim_{M \rightarrow \infty} \frac{1}{M} \left[\frac{1}{2||\bh(0)||^2} + \frac{1}{2||\bh(\frac{M}{2})||^2} \right] + \nonumber\\
&& \quad \lim_{M \rightarrow \infty} \frac{1}{\frac{M}{2}} \sum_{k=1}^{\frac{M}{2}-1} \left[\frac{1}{ \left(||\bh(k)||^2+||\bh(M-k)||^2\right)}\right].\label{Eq:ZFWLLEf}
\end{eqnarray}
For an i.i.d. channel with infinite frequency selectivity, the entries of $\bh(k)$ are i.i.d. complex Gaussian r.v's with zero-mean and unit variance. Therefore, $||\bh(k)||^2$ are i.i.d exponential r.v's with unity variance. Since $||\bh(k)||^2$ always takes positive values, in the limiting case as $M \rightarrow \infty$, the first two terms of $D$ become vanishingly small. Then we end up with
\begin{eqnarray}
D &=& \lim_{M \rightarrow \infty} \frac{1}{\frac{M}{2}} \sum_{k=1}^{\frac{M}{2}-1} \left[\frac{1}{ \left(||\bh(k)||^2+||\bh(M-k)||^2\right)}\right].\label{Eq:ZFWLLEf1}
\end{eqnarray}
The expected value of $D$ is given by
\begin{eqnarray}
E \left[D \right] &=&  \lim_{M \rightarrow \infty} \frac{\frac{M}{2}-1}{\frac{M}{2}}E \left[\frac{1}{\left[||\bh(k)||^2+||\bh(M-k)||^2\right]} \right] \\
&\rightarrow& E \left[\frac{1}{\left[||\bh(k)||^2+||\bh(M-k)||^2\right]} \right].
\end{eqnarray}
Since $\bh(k)$ and $\bh(M-k)$ are i.i.d r.v's, $\left[||\bh(k)||^2+||\bh(M-k)||^2\right]$ is a sum of squares of  $2N_r$ i.i.d. complex Gaussian r.v's. The term $\left[\frac{1}{\left[||\bh(k)||^2+||\bh(M-k)||^2\right]} \right]$ has inverse chi-square distribution with $2N_r$ DOF. Applying the result of \cite{invChisquare} we have: $E \left[\frac{1}{\left[||\bh(k)||^2+||\bh(M-k)||^2\right]} \right]=\frac{1}{2N_r-1}$ and $\texttt{Var} \left[\frac{1}{\left[||\bh(k)||^2+||\bh(M-k)||^2\right]} \right]=\frac{1}{2(2N_r-1)(N_r-1)}$ for $N_r>1$. Therefore, the variance of $D$ is given by
\begin{eqnarray}
\texttt{Var} \left[D \right] &=&  \lim_{M \rightarrow \infty}  \frac{\frac{M}{2}-1}{\frac{M^2}{4}}  \frac{1}{2(2N_r-1)(N_r-1)}\\
&\rightarrow& 0  \quad \texttt{for} \quad N_r>1.
\end{eqnarray}
The post-SNR of WL ZF-LE reaches a constant value of
\begin{eqnarray}
\texttt{SNR}_{\texttt{WL ZF-LE}} &=& \frac{\sigma^{2}_x (2N_r-1)}{\sigma^{2}_n},  \quad \texttt{for} \quad N_r>1 \label{WL ZF-LE snr}.
\end{eqnarray}
Note that the variance of $\left[\frac{1}{\left[||\bh(k)||^2+||\bh(M-k)||^2\right]} \right]$ is bounded only for $N_r>1$.\\
\emph{\textbf{Remark}}\\
In Eq (\ref{Eq:ZFWLLEf}), $||\bh(0)||^2$ and $||\bh(\frac{M}{2})||^2$ are sum of squares of $N_r$ i.i.d. complex Gaussian r.v.'s which gives a chi-square random variable with $N_r$ DOF  while $\left[||\bh(k)||^2+||\bh(M-k)||^2\right]$ has chi-square random variable with $2N_r$ DOF. For the special case of $N_r=1$, the expected value of  $\frac{1}{||\bh(0)||^2}$ or $\frac{1}{||\bh(\frac{M}{2})||^2}$ is unbounded since it has inverse chi-square distribution with one DOF. However, the mean of $\frac{1}{\left[||\bh(k)||^2+||\bh(M-k)||^2\right]}$ is bounded for any value of $N_r$. In the limiting case as $M \rightarrow \infty$, the contribution of the first two terms in (\ref{Eq:ZFWLLEf}) vanishes. However, for the special case of $N_r=1$, and for finite values of $v$, $\bh(k)$ and $\bh(M-k)$ become correlated random variables. Specifically for values of $k=0$ and $k=\frac{M}{2}$, these terms become equal while for values of $k$ in the vicinity of 0 and $\frac{M}{2}$ they become highly correlated. Considering the first two terms of Eq (\ref{Eq:ZFWLLEf}), we see that the terms $\frac{1}{||\bh(0)||^2}$ or $\frac{1}{||\bh(\frac{M}{2})||^2}$ contribute to an increase in the MSE. Similarly, since $\bh(k)$ and $\bh(M-k)$ can be highly correlated for certain subcarrier locations, the term $\frac{1}{\left[||\bh(k)||^2+||\bh(M-k)||^2\right]}$ contributes to an increase in MSE for those subcarrier locations. The overall increase in the MSE can be controlled by considering a WL MMSE which regularizes the denominator terms. Simulation is used to quantify the gain of WL MMSE-LE over ZF case.\\
For detection of real-valued symbols, only the real part of the noise at the output of the equalizer contributes to the error rate. Taking this into account, the post-SNR of a conventional ZF-LE should be modified as
\begin{eqnarray}
\texttt{SNR}_{\texttt{Conv ZF-LE, real}} &=& \frac{2\sigma^{2}_x (N_r-1)}{\sigma^{2}_n} \label{ZF-LE snr, real}
\end{eqnarray}
for $N_r>1$. For $N_r=1$, the WL ZF-LE mitigates both fading and ISI completely and provides a fixed SNR of $\frac{\sigma^{2}_x}{\sigma^{2}_n}$ which is a factor of 2 less than the optimum MFB. However, the conventional ZF-LE performs poorly due to excessive noise enhancement. Joint filtering of signal and its complex-conjugate provides twice the DOF compared to conventional case. These additional DOF aid the equalizer in reducing the noise enhancement and in mitigating fading and ISI effectively. For the case of real-valued modulation, WL equalizers always outperform conventional LEs in multi-path frequency selective channels.
\section{WL MMSE-DFE}\label{wlmmsedfe}
In the WL MMSE-DFE receiver (see Fig \ref{fig:wlmmsedfe}), the received signal and its conjugated-time-reversed replicas are filtered as
\begin{eqnarray*}
\bar{z}(k)&=& \bar{\bw}(k) \bar{\by}(k)
\end{eqnarray*}
where $\bar{\bw}(k)=\left[\bw(k), \bw^{*}(M-k)\right]$ is composed of two vector-valued filters. Next the ISI is eliminated using a feedback filter as
\begin{eqnarray*}
\hat{z}(k)=\bar{z}(k)-\bar{b}(k)x(k).
\end{eqnarray*}
Note that $\bar{b}(k)$ is the frequency domain feedback filter where $1+\bar{b}(k)= 1+\sum_{l=1}^{L} \bar{b}_t(l) e^{\frac{-j 2 \pi kl}{M}}$. The coefficients of the FBF take real-values only. The FBF is implemented in time domain to obtain a decision variable
\begin{eqnarray}
\hat{z}_t(l)= \bar{z}_t(l)-\underbrace{\sum_{m=1}^{L} \bar{b}_t(m) x_t(l \ominus m)}_{\texttt{ISI}}. \label{errort2}
\end{eqnarray}
which is fed to the symbol demodulator. Here, $\bar{z}_t(l)=IDFT[\bar{z}(k)]$ and $\hat{z}_t(l)=IDFT[\hat{z}(k)]$. In Appendix 3, we generalize the MMSE-DFE results for the WL case and provide expressions for the FFF, FBF, and the MSE. We note here that the FFF and FBF expressions are distinct from the ones reported in the literature \cite{Lin}. In addition, the receiver design presented in the Appendix has low implementation complexity. Using the results of Appendix-3, the post-SNR of the WL MMSE-DFE, for $M \rightarrow \infty$, is given by (\ref{wldfepost})
\begin{eqnarray}
\texttt{SNR}_{\texttt{WL MMSE-DFE}} &=& e^{\lim_{M \rightarrow \infty}\frac{1}{M}\sum_{k=0}^{M-1}\ln \left[\frac{\sigma^{2}_x\left(||\bh(k)||^2+||\bh(M-k)||^2\right)+\sigma^{2}_n}{\sigma^{2}_n}\right]}. \label{postsnrWLmmseDFE1}
\end{eqnarray}
\subsection{Performance of WL ZF-DFE in wideband channels }
Setting $\sigma^{2}_n=0$ in (\ref{postsnrWLmmseDFE1}), the post-SNR of a WL ZF-DFE can be expressed as
\begin{eqnarray}
\texttt{SNR}_{\texttt{WL ZF-DFE}} &=& \frac{\sigma^{2}_x}{\sigma^{2}_n} e^{\lim_{M \rightarrow \infty}\frac{1}{M}\sum_{k=0}^{M-1}\ln \left[||\bh(k)||^2+||\bh(M-k)||^2\right]}.\label{exponent}
\end{eqnarray}
Inside the logarithm, we have a sum of  squares of  $2N_r$ i.i.d. complex Gaussian r.v's. In the limiting case as $v \rightarrow \infty$, generalizing the analysis used for the conventional ZF-DFE, we can show that the SNR of WL ZF-DFE reaches a fixed value of
\begin{eqnarray}
\texttt{SNR}_{\texttt{WL ZF-DFE}} &=&  \frac{\sigma^{2}_x e^{\left(-\beta+\sum_{m=1}^{2N_r-1} \frac{1}{m} \right)}}{\sigma^{2}_n}. \label{WL ZF DFE}
\end{eqnarray}
For real-valued modulation, since only the real part of the noise is relevant, the conventional ZF-DFE provides a fixed SNR of
\begin{eqnarray}
\texttt{SNR}_{\texttt{ZF-DFE}} &=&  \frac{2\sigma^{2}_x e^{\left(-\beta+\sum_{m=1}^{N_r-1} \frac{1}{m}\right)}}{\sigma^{2}_n}. \label{ZF DFE snr, real}
\end{eqnarray}
For $N_r=1$, the ideal WL ZF-DFE offers a post-SNR of $\frac{1.5265}{\sigma^{2}_n}$ that is 1.17 dB away from the MFB while the WL ZF-LE has 3.0 dB loss compared to the MFB. The actual performance gap with practical FBF is determined using BER simulation.\\

\emph{\textbf{Remarks}}
\begin{itemize}
	\item For the case of WL ZF/MMSE-DFE, we ignore the potential MSE increase contributed by the terms located at $k=0$ and $k=\frac{M}{2}$. Since at these locations the exponent in (\ref{exponent}) involves the terms $E \left[\ln ||\bh(0)||^2\right]$, $E \left[\ln ||\bh(\frac{M}{2})||^2\right]$ which take a finite value, the overall increase in the MSE can be neglected for finite values of $M$.
\item We note here that our main goal of the paper is to expose the basic properties of conventional and WL equalizers in wideband channels. Our aim is not to promote the use of real constellations over typically used complex-modulations methods. However, the analysis and results related to WL equalizers are useful in systems where real constellations are employed. One such application is discussed in \cite{Kuchi:Nov2012} where binary modulation along with duobinary precoding is employed in the uplink of DFT-precoded-OFDM to reduce the PAPR.
\end{itemize}
\section{Results}\label{results}
We present BER simulation results for BPSK and 8-PSK, and 16-QAM systems. In all cases, the FBF length is set equal to the channel memory. In \cite{Kuchi:2012}, it has already been shown the 20-tap i.i.d. fading channel closely approximates the performance of typical wideband channels such as Ped-B and Veh-A channels. Additionally, it is shown that the BER with a 20-tap i.i.d. channel behaves like an i.i.d. channel with infinite length. Therefore, in the rest of the paper, we present results for a 20-tap i.i.d. Rayleigh fading channel with $M=512$ in all cases.
\subsection{BER results for conventional equalizers}
In the following, we consider the BER performance of the single antenna receivers. In Fig \ref{fig:FIGURE1}, the results are shown for BPSK modulation using ZF receivers. The BER of ideal DFE is close to the conventional DFE up to BER=$10^{-4}$ and shows a degradation at low error rates. This loss is mainly caused by the imperfect initialization of the DFE. Note that the ZF-LE performs poorly due to high noise enhancement. Therefore, initialization of the ZF-DFE with ZE-LE decisions leads to severe error propagation. In Fig \ref{fig:FIGURE2}, BER results are given for 16-QAM system using ZF receivers. The results show that when the ZF-DFE is initialized using known data, the BER follows the ideal DFE case while initialization using the ZF-LE leads to an error floor. In Fig \ref{fig:FIGURE3}, BER is given for 16-QAM employing MMSE based receivers. Unlike ZF case, initialization of the MMSE-DFE using MMSE-LE does not cause severe error propagation and the BER is within 2.0 dB of ideal MMSE-DFE. With lower modulation alphabets, like BPSK, the difference between the BER of MMSE-DFE and ideal MMSE-DFE is small (see Fig \ref{fig:FIGURE4}) and the difference increases for higher order constellations. This loss may be reduced using low-complexity sequence estimation techniques such as reduced state sequence estimation (RSSE) \cite{Eyuboglu:Jan88}. \\
Next, we consider the BER performance of BPSK and 8-PSK system with 2-antennas (See Fig \ref{fig:FIGURE4a}-\ref{fig:FIGURE6}). We observe that in the presence of multiple receiver antennas, the BER of LE improves considerably compared to single antenna case. As a result, initialization of DFE using the LE does not cause significant degradation. Additionally, in dual antenna case, the performance difference between ZF and MMSE based DFEs is found to be negligible.\\
The theoretical performance gap between the post-SNR of the considered receivers and the MFB is tabulated in Table I for an i.i.d. channel with infinite length. In Table II and III we report the gap measured at BERs of 0.01 and 0.001, respectively. For ZF based receivers, the gap measured using simulation in good agreement with the analytically obtained results.
\subsection{BER of WL equalizers}
In Fig \ref{fig:FIGURE5}, we show the BER for WL MMSE based receivers employing BPSK modulation. Comparing with the results of Fig \ref{fig:FIGURE4}, we see that WL processing provides a gain over conventional receivers. In section \ref{WL equalizer}, it is shown that the post-SNR of WL ZF-LE approaches $\frac{(2N_r-1)}{\sigma^2_{n}}$ when $v \rightarrow \infty$. For the case of $N_r=1$ and for finite values of $v$, we argued that a SNR penalty can be expected. For single antenna case, while we expect 3.0 dB SNR gap between the post-SNR of WL ZF-LE and MFB, Table II and III show a gap of 3.2 and 3.8 dB, respectively. However, for the dual antenna case the gap reported in Table II, III is in good agreement with the analytically obtained results given in Table I. Referring to Fig \ref{fig:FIGURE5}-\ref{fig:FIGURE7}, we note that the SNR difference between ideal DFE and actual DFE with LE based initialization is small for both ZF and MMSE cases. The additional DOF obtained by WL processing aid the DFEs to approach ideal performance.
\section{Conclusions}\label{conclusions}
This paper describes the limiting behavior of conventional and WL equalizers in wideband frequency selective channels. For systems employing DFT-precoded-OFDM modulation, closed-form expressions are obtained for the post-SNR of conventional and WL receivers employing ZF-LE and ZF-DFE; simulation is used to assess the performance of MMSE based receivers. In i.i.d. fading channels with infinite channel memory, the post-SNR reaches a fixed value that is comparable to the MFB in most cases. \\

Both conventional MMSE-LE and ZF-LE offer near optimal performance only when the receiver has multiple antennas whereas ideal ZF-DFE and ideal MMSE-DFE perform close to the MFB with a fixed SNR penalty even when the receiver has a single antenna. For single antenna MMSE-DFE with decision feedback, the penalty compared to the ideal DFE is approximately 2.0 dB for 16-QAM systems at high SNRs. The total gap compared to MFB is 4.5 dB. Low-complexity receiver algorithms that further reduce this gap need to be developed. Unlike single antenna case, presence of multiple antennas helps the DFEs to reach a performance close to the MFB.  Multiple receiver antennas are also shown to reduce the error propagation of the DFEs.\\

For single antenna systems employing real-valued modulation alphabets, WL receiver processing can be used to obtain a performance advantage over conventional receivers. In particular, the WL MMSE-LE performs within 3.2-3.8 dB of the MFB while the WL MMSE-DFE reduces the gap with respect to the MFB to 1.0 dB. Results show that the multi-antenna WL receivers (both LEs and DFEs) perform very close to the MFB as predicted by the infinite length i.i.d. fading channel model.\\

We note here that the proposed infinite length i.i.d fading channel model can be used to obtain the limiting performance of MIMO systems employing spatial multiplexing (SM). The analysis has been carried out in \cite{Kuchi:2012a} for the case of MIMO ZF-LE where it is shown that the post-SNR of the receiver reaches a constant value of $\frac{N_r-N_t}{\sigma^2_{n}}$ for $N_r>N_t$, where $N_t$ is the SM rate. Extension to the general case of SM employing ZF/MMSE-DFEs is yet to be considered.
\section{Appendix}
\subsection{Appendix-1: Derivation of MMSE-DFE filter settings}
We obtain closed-form expressions for the FFF, FBF and MSE for the case when the FBF is restricted to have finite length. First, we show that the MSE minimizing solution for the FBF becomes a finite length prediction error filter that whitens the error covariance at the output of the  MMSE-LE. The solution obtained in our case becomes a multiple receiver antenna generalization of the results presented in \cite{Italia:02}.  Similarly, \cite{Gerstacker:may08} presented an alternative approach for multiple-input-multiple-output (MIMO) systems where the problem of designing the FBF is formulated as one of finite length prediction error filter design. This alternative approach results in a solution that agrees with our results for the multiple receiver antennas case. We further generalize our results to the infinite length filter case which facilitates performance analysis in i.i.d. fading channels. The derivations presented in this section follow the approach presented in \cite{Cioffi:Stanford}, \cite{AlDhahir:July95}.
We define an error signal
\begin{eqnarray}
e(k)=\vec{z}(k)-x(k). \label{error}
\end{eqnarray}
This is written in time domain as
\begin{eqnarray}
e_t(l)= z_t(l)- x_t(l )-\sum_{m=1}^{L} b_t(m) x_t(l \ominus m). \label{errort}
\end{eqnarray}
where the symbol $\ominus$ denotes left circular shift operation. Define: $r_{ee}(k)=E\left(||e(k)||^2\right)$. Using Parseval's theorem: $\frac{1}{M} \sum_{k=0}^{M-1} ||e(k)||^2=\sum_{l=0}^{M-1} ||e_t(l)||^2$. Taking expectation on both sides, we get
\begin{eqnarray*}
\frac{1}{M} \sum_{k=0}^{M-1} r_{ee}(k)= \sum_{l=0}^{M} E\left(||e_t(l)||^2\right).
\end{eqnarray*}
The MSE is defined as: $\texttt{MSE}=E\left(||e_t(l)||^2\right)$ which is independent of the time index $l$. It can be written as
\begin{eqnarray*}
\texttt{MSE}=\frac{1}{M^2} \sum_{k=0}^{M-1} r_{ee}(k).
\end{eqnarray*}
Next, we obtain an expression for the FFF in frequency domain. Applying orthogonality principle \cite{Cioffi:Stanford}
\begin{eqnarray}
E\left(e(k)\by^{\dagger}(k)\right)=0, \quad \texttt{for} \texttt{\quad} k=0,1,..,M-1.   \label{mmsedfe ortho}
\end{eqnarray}
Substituting  (\ref{error}), in (\ref{mmsedfe ortho}), and evaluating the expectation, the FFF can be expressed as
\begin{eqnarray}
\bw(k) &=& (1+b(k))\mathbf{R}_{xy}(k) \mathbf{R}^{-1}_{yy}(k)
\end{eqnarray}
where $\mathbf{R}_{xy}(k)=E\left(x(k)\by^{\dagger}(k)\right)=r_{xx}(k) \bh^{\dagger}(k)$ and $\mathbf{R}_{yy}(k)=E\left(\by(k)\by^{\dagger}(k)\right)=\left[ \bh(k)r_{xx}(k)\bh^{\dagger}(k) + \mathbf{R}_{nn}(k) \right]$. Here $r_{xx}(k)=E\left(||x(k)||^2\right)=M \sigma^{2}_x$ and $\mathbf{R}_{nn}(k)=E\left(\bn(k) \bn^{\dagger}(k)\right)=M \sigma^{2}_n \bI$.
The FFF can be expressed in alternative form as
\begin{eqnarray}
\bw(k) &=& (1+b(k)) r_{xx}(k) \bh^{\dagger}(k) \left[ \bh(k) r_{xx}(k)\bh^{\dagger}(k) + \mathbf{R}_{nn}(k) \right]^{-1}\\
       &=& (1+b(k))  \left[ \mathbf{R}^{-1}_{xx}(k)+\bh^{\dagger}(k) \mathbf{R}^{-1}_{nn}(k)\bh(k) \right]^{-1} \bh^{\dagger}(k) \mathbf{R}^{-1}_{nn}(k)\\ \label{lemma}
       &=& \frac{(1+b(k)) }{|\bh(k)|^2+\frac{\sigma^{2}_n}{\sigma^{2}_x}}\bh^{\dagger}(k).\label{FFFconv}
\end{eqnarray}
Note that (\ref{lemma}) follows from applying matrix inversion lemma\footnote{$\left(A+BCD\right)^{-1}=A^{-1}+A^{-1}B\left(C^{-1}+DA^{-1}B\right)^{-1}DA^{-1}$}. With this  choice of FFF, the minimum MSE can be shown to be \cite{Cioffi:Stanford}, \cite{AlDhahir:July95}
\begin{eqnarray} \label{msec}
\texttt{MSE}= \frac{1}{M^2}\sum_{k=0}^{M-1} r_{ee}(k)= \frac{1}{M}\sum_{k=0}^{M-1} \frac{\sigma^{2}_n |(1+b(k))|^2}{||\bh(k)||^2+\frac{\sigma^{2}_n}{\sigma^{2}_x}}.
\end{eqnarray}
To obtain the coefficients $b_t(l)$, we take the partial derivatives
\begin{eqnarray}
\frac{\partial \texttt{MSE }}{\partial b_t(l)}= \frac{1}{M}\sum_{k=0}^{M-1} \frac{\sigma^{2}_n \frac{\partial }{\partial b_t(l)} \left[||(1+\sum_{l=1}^{L} b_t(l) e^{-\frac{j 2 \pi kl}{M}})||^2\right]}{||\bh(k)||^2+\frac{\sigma^{2}_n}{\sigma^{2}_x}}  \label{partial}
\end{eqnarray}
Let
\begin{eqnarray*}
\frac{\partial }{\partial b_t(l)} \left[||(1+\sum_{l=1}^{L} b_t(l) e^{-\frac{j 2 \pi kl}{M}})||^2\right]&=&\frac{\partial }{\partial b_t(l)} \left[(1+\sum_{l=1}^{L} b_t(l) e^{-\frac{j 2 \pi kl}{M}}+\sum_{l=1}^{L} b^{*}_t(l) e^{\frac{j 2 \pi kl}{M}}+ \sum_{l=1}^{L} \sum_{m=1}^{L}b_t(l) b^{*}_t(m) e^{\frac{-j 2 \pi k(l-m)}{M}})\right].
\end{eqnarray*}
Treating $b_t(l)$ and $b^{*}_t(l)$ as independent variables we get
$
\frac{\partial }{\partial b_t(l)} \left[||(1+\sum_{l=1}^{L} b_t(l) e^{-\frac{j 2 \pi kl}{M}})||^2\right] =  \left[ e^{-\frac{j 2 \pi kl}{M}} + \sum_{m=1}^{L} b^{*}_t(m) e^{\frac{-j 2 \pi k(l-m)}{M}})\right].
$
Substituting this result in (\ref{partial}) and setting the partial derivatives to zero, we obtain the MSE minimizing condition
\begin{eqnarray}
\frac{1}{M}\sum_{k=0}^{M-1} \frac{\sigma^{2}_n \left[ e^{-\frac{j 2 \pi kl}{M}} + \sum_{m=1}^{L} b^{*}_t(m) e^{\frac{-j 2 \pi k(l-m)}{M}})\right]}{||\bh(k)||^2+\frac{\sigma^{2}_n}{\sigma^{2}_x}}=0. \label{sol}
\end{eqnarray}
We define the following IDFT pair
\begin{eqnarray}\label{ql}
q(l)=\frac{1}{M}\sum_{k=0}^{M-1} \frac{\sigma^{2}_n}{||\bh(k)||^2+\frac{\sigma^{2}_n}{\sigma^{2}_x}} e^{j 2 \pi k l}, \quad l=0,1,..,M-1.
\end{eqnarray}
Note that$ \frac{\sigma^{2}_n}{||\bh(k)||^2+\frac{\sigma^{2}_n}{\sigma^{2}_x}}$ is the frequency domain error covariance at the output of a MMSE-LE \cite{Cioffi:Stanford} where $q(l)$ is the corresponding time domain error covariance.  Now, we can express (\ref{sol}) in compact form as
\begin{eqnarray}\label{keyeq}
\mathbf{A} \mathbf{b}^{*} = -\mathbf{q}^{*}
\end{eqnarray}
where the $(l,m)$th element of the matrix $\mathbf{A}$ is given by: $A(l,m)=q(m-l)$, $\mathbf{b}=[b_t(1), b_t(2),..,b_t(L)]^{Tr}$, and $\mathbf{q}=[q(1), q(2),..,q(L+1)]^{Tr}$. The elements of the FBF can be obtained by solving (\ref{keyeq}). It can be seen that the MSE minimizing solution for the FBF becomes a finite length prediction error filter of order $L$ that whitens the error covariance at the output of the  MMSE-LE. The FBF coefficients can be calculated efficiently using the Levinson-Durbin recursion. The minimum MSE can be obtained by substituting the values of the FBF coefficients in the MSE expression (\ref{msec}). \\
Next we characterize the MMSE-DFE for the case of $M \rightarrow \infty$. \\
Expanding the log spectrum using the DFT
\begin{eqnarray}
\ln \left[||\bh(k)||^2+\frac{\sigma^{2}_n}{\sigma^{2}_x} \right]=  \sum_{l=0}^{M-1} c(l) e^{\frac{-j2 \pi l k}{M}}, \quad k=0,1,..,M-1 \label{logsf}
\end{eqnarray}
where
\begin{eqnarray}
 c(l)= \frac{1}{M } \sum_{k=0}^{M-1}\ln \left[||\bh(k)||^2+\frac{\sigma^{2}_n}{\sigma^{2}_x} \right]  e^{\frac{j2 \pi l k}{M}}, \quad l=0,2,..,M-1 \label{logsf1}
\end{eqnarray}
with $c(0)=\frac{1}{M } \sum_{k=0}^{M-1}\ln \left[||\bh(k)||^2+\frac{\sigma^{2}_n}{\sigma^{2}_x} \right]$ and $c(l)=c^{*}(-l)=c^{*}(M \ominus l)$.
For odd values of $M$, let us rewrite (\ref{logsf}) as
\begin{eqnarray*}
\left[||\bh(k)||^2+\frac{\sigma^{2}_n}{\sigma^{2}_x} \right] &=&  e^{\sum_{l=0}^{M-1} c(l) e^{\frac{-j2 \pi l k}{M}} }\\
&=& e^{c_0} e^{\sum_{l=1}^{\frac{M-1}{2}} c(l) e^{-\frac{j2 \pi l k}{M}}} e^{\sum_{l=\frac{M-1}{2}+1}^{M-1} c^*(M-l) e^{\frac{-j2 \pi l k}{M}}}\\
&=& e^{c_0} e^{\sum_{l=1}^{\frac{M-1}{2}} c(l) e^{-\frac{j2 \pi l k}{M}}} e^{\sum_{l=1}^{\frac{M-1}{2}} c^*(l) e^{\frac{-j2 \pi l k}{M}}}\\
&=& \gamma g(k) g^{*}(k) \label{sf}
\end{eqnarray*}
where $\gamma =e^{c(0)}$, $g(k)=e^{\sum_{l=1}^{\frac{M-1}{2}} c(l) e^{\frac{-j2 \pi l k}{M}}}$. In the limiting case as $M \rightarrow \infty$, the DFTs approaches discrete-time-Fourier-transforms (DTFTs) i.e.,
\begin{eqnarray*}
\bh(k) \rightarrow \bh(f), \quad g(k) \rightarrow g(f)
\end{eqnarray*}
where
\begin{eqnarray*}
g(f) &=& e^{\sum_{l=1}^{\infty} c(l) e^{-j2 \pi l f}}\\
     &=&  1+\sum_{q=1}^{\infty}\frac{\left(\sum_{l=1}^{\infty} c(l) e^{-j2 \pi l f}\right)^q}{q}\\
     &=& 1+ \sum_{l=1}^{\infty} g_t(l)  e^{-j2 \pi l f}
\end{eqnarray*}
The result on line-2 is obtained by expanding the exponential function into an infinite series. The coefficients $g_t(l)$ are obtained by collecting appropriate terms in the summation on line-2.  Note that $g(f)$ is a causal, monic, and minimum-phase filter with all its poles and zeros inside the unit circle. Similarly, $g^{*}(f)$ is a non-causal, monic, and maximum phase filter. For $M \rightarrow \infty$, (\ref{logsf1}) can be written as
\begin{eqnarray*}
c(l) &=& \lim_{M \rightarrow \infty} \frac{1}{M } \sum_{k=0}^{M-1}\ln\left[||\bh(k)||^2+\frac{\sigma^{2}_n}{\sigma^{2}_x} \right]  e^{\frac{j2 \pi l k}{M}}\\
     &\rightarrow& \int_{0}^{1} \ln \left[||\bh(f)||^2+\frac{\sigma^{2}_n}{\sigma^{2}_x} \right]  e^{j2\pi l f} df
\end{eqnarray*}
where $\bh(f)=  \sum_{l=0}^{\infty} \bh_t(l) e^{j2\pi l f}$ is the DTFT of $\bh_t(l)$ which is a periodic in $f$ with period 1. Further,
\begin{eqnarray*}
\gamma &=& e^{\lim_{M \rightarrow \infty} \frac{1}{M } \sum_{k=0}^{M-1}\ln\left[||\bh(k)||^2+\frac{\sigma^{2}_n}{\sigma^{2}_x} \right] }\\
&\rightarrow& e^{ \int_{0}^{1} \ln\left[||\bh(f)||^2+\frac{\sigma^{2}_n}{\sigma^{2}_x} \right] df}.
\end{eqnarray*}
Now we write the spectrum factorization for the continuous case as \cite{Cioffi:Stanford}, \cite{AlDhahir:July95}
\begin{eqnarray}
\left[||\bh(f)||^2+\frac{\sigma^{2}_n}{\sigma^{2}_x} \right] &=& \gamma g(f) g^{*}(f). \label{scalarsf}
\end{eqnarray}
The MSE given by (\ref{msec}) becomes
\begin{eqnarray}
\texttt{MSE} &=& \lim_{M \rightarrow \infty} \frac{1}{M}\sum_{k=0}^{M-1} \frac{\sigma^{2}_n ||1+b(k)||^2}{||\bh(k)||^2+\frac{\sigma^{2}_n}{\sigma^{2}_x}}\\ \label{cmsecc}
&\rightarrow& \int_{0}^{1} \frac{\sigma^{2}_n ||1+b(f)||^2}{||\bh(f)||^2+\frac{\sigma^{2}_n}{\sigma^{2}_x}} df \label{cmsecc1}
\end{eqnarray}
where
\begin{eqnarray*}
1+b(f) &=& 1+ \sum_{l=1}^{\infty} b_t(l) e^{-j2 \pi l f}.
\end{eqnarray*}
The optimum choice which minimizes the MSE given in (\ref{cmsecc1}) is given by: $1+b(f)=g(f)$ \cite{Cioffi:Stanford}, \cite{AlDhahir:July95}. Using this, and substituting (\ref{scalarsf}) in the MSE expression (\ref{cmsecc1}) we obtain the minimum MSE as
\begin{eqnarray*}
\texttt{MSE}_{M \rightarrow \infty}  &=& \frac{\sigma^{2}_n}{\gamma}.
\end{eqnarray*}
Assuming ideal decision feedback, the SNR at the output of the MMSE-DFE is given by
\begin{eqnarray*}
\texttt{SNR}_{\texttt{MMSE-DFE}} &=& \frac{\sigma^{2}_x}{\texttt{MSE}}\\
&=& \frac{\sigma^{2}_x}{\sigma^{2}_n} e^{\lim_{M \rightarrow \infty}\frac{1}{M}\sum_{k=0}^{M-1}\ln \left[\bh^{\dagger}(k)\bh(k)+\frac{\sigma^{2}_n}{\sigma^{2}_x}\right]}\\
&=& e^{\lim_{M \rightarrow \infty}\frac{1}{M}\sum_{k=0}^{M-1}\ln \left[\frac{\sigma^{2}_x}{\sigma^{2}_n}||\bh(k)||^2+1\right]}\\
&\rightarrow& e^{\int_{0}^{1} \ln \left[\frac{\sigma^{2}_x}{\sigma^{2}_n}||\bh(f)||^2+1\right] df}.
\end{eqnarray*}
\subsection{Appendix-3: Derivation of WL MMSE-DFE filter settings}
In this section we discuss the design aspects of WL MMSE DFE. The key implementation differences between conventional and WL equalizers are highlighted. Specifically, we notice that the noise covariance term at the output of the WL MMSE section exhibits even symmetry in frequency domain. This property is exploited to reduce the computational complexity of FFF and FBF filter calculation.\\
The error signal for WL case is defined as
\begin{eqnarray*}
\bar{e}(k)=\hat{z}(k)-x(k).
\end{eqnarray*}
In time domain
\begin{eqnarray*}
\bar{e}_t(l)= \bar{z}_t(l)- \sum_{m=0}^{L} \bar{b}_t(m)x_t(l \ominus m)
\end{eqnarray*}
where $\bar{b}_t(l)=IDFT(\bar{b}(k))$.
Let $\bar{r}_{ee}(k)=E(||\bar{e}(k)||^2)$. The total MSE is given by
\begin{eqnarray*}
\texttt{MSE}_{\texttt{WL DFE}}= \frac{1}{M^2}\sum_{k=0}^{M-1} \bar{r}_{ee}(k)= \frac{1}{M^2}\sum_{k=0}^{M-1} E(||\bar{e}(k)||^2).
\end{eqnarray*}
Following the conventional MMSE-DFE case, the MSE minimizing solution for the WL FFF can be obtained as \cite{Cioffi:Stanford},\cite{AlDhahir:July95}
\begin{eqnarray*}
\bar{\bw}(k)  &=& \frac{1+\bar{b}(k)}{||\bh(k)||^2+||\bh(M-k)||^2+\frac{\sigma^{2}_n}{\sigma^{2}_x}}\bar{\bh}^{\dagger}(k)\\
              &=& \frac{1+\bar{b}(k)}{||\bh(k)||^2+||\bh(M-k)||^2+\frac{\sigma^{2}_n}{\sigma^{2}_x}} \left[\bh^{*}(k), \bh(M-k) \right].
\end{eqnarray*}
Let $P(k)=||\bh(k)||^2+||\bh(M-k)||^2+\frac{\sigma^{2}_n}{\sigma^{2}_x}$. This is a real-valued function which exhibits even symmetry i.e., $P(k)=P(M-k)$ for $k=0,1,..,M-1$. Similar to the WL MMSE-LE case, it is computationally efficient to calculate the filter  $\bw(k)$ explicitly. The second filter can be obtained from the first by applying complex-conjugation and frequency reversal operations.\\
The minimum MSE can be expressed as \cite{Cioffi:Stanford},\cite{AlDhahir:July95}
\begin{eqnarray*}
\texttt{MSE}_{\texttt{WL DFE}}= \frac{1}{M}\sum_{k=0}^{M-1} \frac{\sigma^{2}_n ||1+\bar{b}(k)||^2}{P(k)}.
\end{eqnarray*}
Consider the partial derivatives
\begin{eqnarray}
\frac{\partial \texttt{MSE}_{\texttt{WL DFE}} }{\partial \bar{b}_t(l)}= \frac{1}{M}\sum_{k=0}^{M-1} \frac{M \sigma^{2}_n \frac{\partial }{\partial \bar{b}_t(l)} \left[||(1+\sum_{l=1}^{L} \bar{b}_t(l) e^{-\frac{j 2 \pi kl}{M}})||^2\right]}{P(k)}.\label{partialderwl}
\end{eqnarray}
Let
\begin{eqnarray*}
\frac{\partial }{\partial \bar{b}_t(l)} \left[||(1+\sum_{l=1}^{L} \bar{b}_t(l) e^{-\frac{j 2 \pi kl}{M}})||^2\right]&=&\frac{\partial }{\partial \bar{b}_t(l)} \left[(1+\sum_{l=1}^{L} 2 \bar{b}_t(l) \cos \left(\frac{ 2 \pi kl}{M} \right)+ \sum_{l=1}^{L} \sum_{m=1}^{L} \bar{b}_t(l) \bar{b}_t(m) e^{\frac{-j 2 \pi k(l-m)}{M}})\right]\\
&=&  2\cos \left(\frac{2 \pi kl}{M} \right) + \sum_{m=1}^{L} 2 \bar{b}_t(m) \cos \left(\frac{ 2 \pi k(l-m)}{M} \right).
\end{eqnarray*}
Substituting this result in (\ref{partialderwl}) and setting the partial derivatives to zero, we get
\begin{eqnarray}
\frac{1}{M}\sum_{k=0}^{M-1} \frac{M \sigma^{2}_n \left[2 \cos \left(\frac{ 2 \pi kl}{M} \right) + \sum_{m=1}^{L} 2 b_t(m)\cos \left(\frac{ 2 \pi k(l-m)}{M} \right) \right]}{P(k)}=0. \label{realFBF}
\end{eqnarray}
Let us define the following transform pair
\begin{eqnarray}\label{wlql}
\bar{q}(l)=\frac{1}{M}\sum_{k=0}^{M-1} \frac{\sigma^{2}_n}{P(k)} \cos \left(\frac{2 \pi kl}{M} \right).
\end{eqnarray}
It can be implemented with low-complexity using standard IFFT algorithm. Alternatively, noting that $P(k)=P(M-k)$, we can write (\ref{wlql}) as
\begin{eqnarray}
\bar{q}(l) &=& \frac{1}{M}\left[ P(0)+P\left(\frac{M}{2}\right) + \sum_{k=0}^{\frac{M}{2}-1} P(k)  \cos \left(\frac{2 \pi kl}{M} \right) + \right. \nonumber \\
&&\left.\sum_{k=\frac{M}{2}+1}^{M-1} P(M-k) \cos \left(\frac{2 \pi kl}{M} \right)  \right]\\
&=& \frac{2}{M}\left[ \frac{\left(P(0)+P\left(\frac{M}{2}\right)\right)}{2} + \sum_{k=0}^{\frac{M}{2}-1} P(k)  \cos \left(\frac{2 \pi kl}{M} \right)  \right], \\
&& \quad l=0,1,..,\frac{M}{2}.\nonumber
\end{eqnarray}
The last term involves $\frac{M}{2}$  point type-1 DCT of $P(k)$. Note that $\bar{q}(l)$ needs to be calculated only for the first $\frac{M}{2}$ terms since the rest of the coefficients can be obtained exploiting the even symmetry of $\bar{q}(l)$ i.e., $\bar{q}(l)=\bar{q}(M-l)$.\\
The MSE minimizing condition (\ref{realFBF}) can be expressed in vector-matrix form as
\begin{eqnarray}\label{Syseq}
\bar{\mathbf{A}} \bar{\mathbf{b}} =-\bar{\mathbf{q}}
\end{eqnarray}
where the $(l,m)$th element of the matrix $\bar{\mathbf{A}}$ denotes as $\bar{A}(l,m)$ is given by: $\bar{A}(l,m)=\bar{q}(m-l)$, $\bar{\mathbf{b}}=[\bar{b}_t(1), \bar{b}_t(2),..,\bar{b}_t(L)]^{Tr}$, and $\bar{\mathbf{q}}=[\bar{q}(1), \bar{q}(2),..,\bar{q}(L)]^{Tr}$. Note that FBF can be calculated with low-complexity using Levinson-Durbin recursion which involves real-valued quantities whereas the FBF for the conventional case involves complex values. Now we consider the infinite length filter case.\\
Let us consider
\begin{eqnarray*}
\ln \left[||\bh(k)||^2+||\bh(M-k)||^2+\frac{\sigma^{2}_n}{\sigma^{2}_x} \right]=  \sum_{l=0}^{M-1} \bar{c}(l) e^{\frac{-j2 \pi l k}{M}}, \quad k=0,1,..,M-1
\end{eqnarray*}
where the real-valued coefficients $ \bar{c}(l)$ can be obtained using the IDFT,
\begin{eqnarray*}
 \bar{c}(l)= \frac{1}{M } \sum_{k=0}^{M-1}\ln \left[||\bh(k)||^2+||\bh(M-k)||^2+\frac{\sigma^{2}_n}{\sigma^{2}_x} \right]  e^{\frac{j2 \pi l k}{M}}
\end{eqnarray*}
with $\bar{c}(0)=\frac{1}{M } \sum_{k=0}^{M-1}\ln \left[||\bh(k)||^2+||\bh(M-k)||^2+\frac{\sigma^{2}_n}{\sigma^{2}_x} \right]$ and $\bar{c}(l)=\bar{c}(M-l)$.
Let us consider
\begin{eqnarray*}
\left[||\bh(k)||^2+||\bh(M-k)||^2+\frac{\sigma^{2}_n}{\sigma^{2}_x} \right] &=&  e^{\sum_{l=0}^{M-1} \bar{c}(l) e^{\frac{-j2 \pi l k}{M}} }\\
&=& e^{\bar{c}(0)} e^{\sum_{l=1}^{\frac{M-1}{2}} \bar{c}(l) e^{\frac{-j2 \pi l k}{M}}} e^{\sum_{l=\frac{M-1}{2}+1}^{M-1} \bar{c}(l) e^{\frac{-j2 \pi l k}{M}}}\\
&=& \bar{\gamma} \bar{g}(k) \bar{g}^{*}(k)
\end{eqnarray*}
where $\bar{\gamma} =e^{\bar{c}(0)}$, $\bar{g}(k)=e^{\sum_{l=1}^{\frac{M-1}{2}} \bar{c}(l) e^{\frac{-j2 \pi l k}{M}}}$.\\
As $M \rightarrow \infty$, $\bar{g}(k) \rightarrow \bar{g}(f)$ where
\begin{eqnarray*}
\bar{g}(f) &=& e^{\sum_{l=1}^{\infty} \bar{c}(l) e^{-j2 \pi l f}}\\
     &=& 1+ \sum_{l=1}^{\infty} \bar{g}_t(l)  e^{-j2 \pi l f}.
\end{eqnarray*}
Further, we have
\begin{eqnarray*}
\bar{c}(l) &=& \lim_{M \rightarrow \infty} \frac{1}{M } \sum_{k=0}^{M-1}\ln \left[||\bh(k)||^2+||\bh(M-k)||^2+\frac{\sigma^{2}_n}{\sigma^{2}_x} \right] e^{\frac{j2 \pi l k}{M}}\\
     &\rightarrow& \int_{0}^{1} \ln\left[||\bh(f)||^2+||\bh(-f)||^2+\frac{\sigma^{2}_n}{\sigma^{2}_x} \right]  e^{j2\pi l f} df.
\end{eqnarray*}
Next,
\begin{eqnarray*}
\bar{\gamma} &=& e^{ \lim_{M \rightarrow \infty} \frac{1}{M } \sum_{k=0}^{M-1}\ln \left[||\bh(k)||^2+||\bh^{*}(M-k)||^2+\frac{\sigma^{2}_n}{\sigma^{2}_x} \right] }\\
 &\rightarrow& e^{ \int_{0}^{1} \ln \left[||\bh(f)||^2+||\bh(-f)||^2+\frac{\sigma^{2}_n}{\sigma^{2}_x} \right] df }.
\end{eqnarray*}
We write the spectrum factorization for this case
\begin{eqnarray*}
\left[||\bh(f)||^2+||\bh(-f)||^2+\frac{\sigma^{2}_n}{\sigma^{2}_x} \right] &=& \bar{\gamma} \bar{g}(f) \bar{g}^{\dagger}(f).
\end{eqnarray*}
The MSE for this case becomes
\begin{eqnarray*}
\texttt{MSE}_{\texttt{WL MMSE-DFE}} &=&  \lim_{M \rightarrow \infty} \frac{1}{M}\sum_{k=0}^{M-1} \frac{\sigma^{2}_n ||1+\bar{b}(k)||^2}{||\bh(k)||^2+||\bh(M-k)||^2+\frac{\sigma^{2}_n}{\sigma^{2}_x}}\\
&\rightarrow& \int_{0}^{1} \frac{\sigma^{2}_n ||1+\bar{b}(f)||^2}{||\bh(f)||^2+||\bh(-f)||^2+\frac{\sigma^{2}_n}{\sigma^{2}_x}} df.
\end{eqnarray*}
The MSE is minimized with the choice: $1+\bar{b}(f)=\bar{g}(f)$. Using this, we obtain the minimum MSE as
\begin{eqnarray*}
\texttt{MSE}_{M \rightarrow \infty}  &=& \frac{\sigma^{2}_n}{\bar{\gamma}}.
\end{eqnarray*}
Assuming ideal decision feedback, the SNR of the WL MMSE-DFE is given by
\begin{eqnarray}
\texttt{SNR}_{\texttt{WL MMSE-DFE}} &=& \frac{\sigma^{2}_x}{\texttt{MSE}_{\texttt{WL MMSE DFE}}}\nonumber\\
&=& \frac{\sigma^{2}_x}{\sigma^{2}_n} e^{\lim_{M \rightarrow \infty}\frac{1}{M}\sum_{k=0}^{M-1}\ln \left[||\bh(k)||^2+||\bh(M-k)||^2+\frac{\sigma^{2}_n}{\sigma^{2}_x}\right]}\nonumber\\
&=& e^{\lim_{M \rightarrow \infty}\frac{1}{M}\sum_{k=0}^{M-1}\ln \left[\frac{\sigma^{2}_x}{\sigma^{2}_n}\left(||\bh(k)||^2+||\bh(M-k)||^2\right)+1\right]}\label{wldfepost}\\
&\rightarrow & e^{\int_{0}^{1} \ln \left[\frac{\sigma^{2}_x}{\sigma^{2}_n}\left(||\bh(f)||^2+||\bh(-f)||^2\right)+1\right] df}.\nonumber
\end{eqnarray}\\

\bibliographystyle{IEEEtran} 
\bibliography{IEEEabrv,IEEEexample}

\tikzstyle{block} = [draw,  rectangle, minimum height=3em, minimum width=6em]
\tikzstyle{sum} = [draw, circle, node distance=1cm]
\tikzstyle{input} = [coordinate]
\tikzstyle{output} = [coordinate]
\tikzstyle{pinstyle} = [pin edge={to-,thin,black}]

\begin{figure}[h!]
\scalebox{0.9}{
\begin{tikzpicture}[auto, node distance=2cm,>=latex']
 \node [input, name=input] {};
 \node [block, right of=input] (wk) {$\mathbf{w}(k)$};
 \node [block, right of=wk,node distance=3cm] (idft) {IDFT};\draw [->] (wk) -- node[name=u] {} (idft);
 \node [sum, right of=idft,node distance=2.5cm] (sum) {};
\node [block, right of=sum,node distance=4cm] (dem) {Demod};

\node [output, right of=dem,node distance=2.5cm] (output) {};

\draw [draw,->] (input) -- node {$\mathbf{y}(k)$} (wk);
\draw [->] (idft) -- node {$z_t(l)$} (sum);
\draw [->] (sum) -- node[name=s2d] {} (dem);
\draw [->] (dem) -- node [name=y] {$x_t(l)$}(output);
\node [block, below of=s2d] (fbf)at (9.5,0) {\begin{tabular}{c}FBF\\ $1-b_t(l)$ \end{tabular}} ;
\draw [->] (dem) |- (fbf);
\draw [->] (fbf) -| node[pos=0.99] {$-$}
        node [near end] {} (sum);
\end{tikzpicture}}
\caption{Conventional MMSE-DFE} \label{fig:convmmsedfe}
\end{figure}
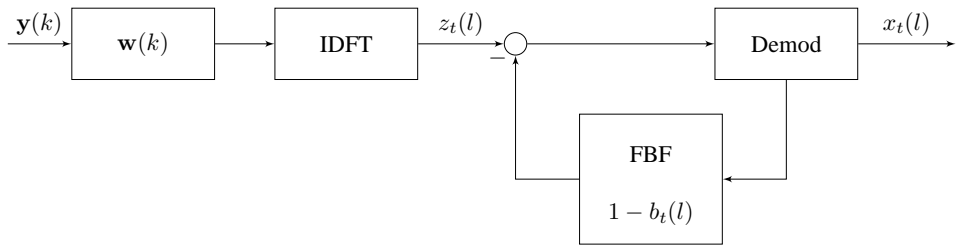


\tikzstyle{block} = [draw,  rectangle, minimum height=3em, minimum width=6em]
\tikzstyle{sum} = [draw, circle, node distance=1cm]
\tikzstyle{output} = [coordinate]
\begin{figure}[h!]
\scalebox{0.9}{
\begin{tikzpicture}[auto, node distance=2cm,>=latex']
\node at (0,-0.7) (input2) {$\mathbf{y}(k)$};
\node at (0,-1) (input1) {};
\node at (0.3,-2.7) (input2) {$\mathbf{y}^\ast(M-k)$};
\node at (0,-3) (input2) {};
\node [block, right of=input1,node distance=2.7cm] (wk) {$\mathbf{w}(k)$};
\node [block, right of=input2,node distance=2.7cm] (wsk) {$\mathbf{w}^\ast(M-k)$};
\node at (5,-1) (input4) {};
\node at (5,-2) (input3) {$+$};
\node at (5,-3) (input5) {};
\node [sum, right of=input3,node distance=0cm] (sum) {};
\draw[->] (input1) -- (wk);
\draw[->] (input2) -- (wsk);
\draw[-] (wk) -- (5,-1);
\draw[-] (wsk) -- (5,-3);
\draw[->] (5,-1) -- (sum);
\draw[->] (5,-3) -- (sum);

\node [block, right of=input3,node distance=2.3cm] (idft) {IDFT};
\draw[->] (sum) -- (idft);
\draw [draw,->] (sum) -- node {$\bar{z}(k)$} (idft);
\node at (9,-2) (output) {};
\draw[->] (idft) -- (9,-2);\node at (9.1,-1.7) (output) {$x_t(l)$};
\end{tikzpicture}}
\caption{WL MMSE-LE} \label{fig:wlmmsele}
\end{figure}
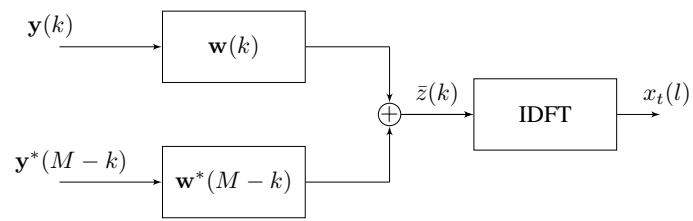

\begin{figure}[h!]
\scalebox{0.9}{
\begin{tikzpicture}[auto, node distance=2cm,>=latex']
\node at (0,-0.7) (input2) {$\mathbf{y}(k)$};
\node at (0,-1) (input1) {};
\node at (0.3,-2.7) (input2) {$\mathbf{y}^\ast(M-k)$};
\node at (0,-3) (input2) {};
\node [block, right of=input1,node distance=2.7cm] (wk) {$\mathbf{w}(k)$};
\node [block, right of=input2,node distance=2.7cm] (wsk) {$\mathbf{w}^\ast(M-k)$};
\node at (5,-1) (input4) {};
\node at (5,-2) (input3) {$+$};
\node at (5,-3) (input5) {};
\node [sum, right of=input3,node distance=0cm] (sum) {};
\draw[->] (input1) -- (wk);
\draw[->] (input2) -- (wsk);
\draw[-] (wk) -- (5,-1);
\draw[-] (wsk) -- (5,-3);
\draw[->] (5,-1) -- (sum);
\draw[->] (5,-3) -- (sum);

\node [block, right of=input3,node distance=2.3cm] (idft) {IDFT};
\draw[->] (sum) -- (idft);
\draw [draw,->] (sum) -- node {$\bar{\mathbf{z}}(k)$} (idft);
\node at (9.5,-2) (output) {};
\draw[-] (idft) -- (output);
\node at (9.1,-1.7) (output1) {$\bar{z}_t(l)$};
\node [sum, right of=output,node distance=0.2cm] (sum2) {};
\node at (9.7,-2) (input6) {$+$};
\draw[->] (output) -- (sum2);
\node at (9.7,-3.8) (input7) {};
\node [block, right of=input7,node distance=2cm] (fbf) {\begin{tabular}{c}FBF\\ $1-b_t(l)$ \end{tabular}};
\draw[-] (fbf) -- (9.7,-3.8);
\draw[->] (9.7,-3.8) -- (sum2);
\node [block, right of=input6,node distance=3.9cm] (dem) {Demod};
\draw[->] (sum2) -- (dem);
\node at (13.6,-3.8) (input8) {};
\draw[-] (dem) -- (13.6,-3.8);
\draw[->] (13.6,-3.8) -- (fbf);
\node at (16,-2) (outputf) {};
\draw[->] (dem) -- (outputf);
\node at (15.5,-1.7) (output1) {$x_t(l)$};
\end{tikzpicture}}
\caption{WL MMSE-DFE} \label{fig:wlmmsedfe}
\end{figure}
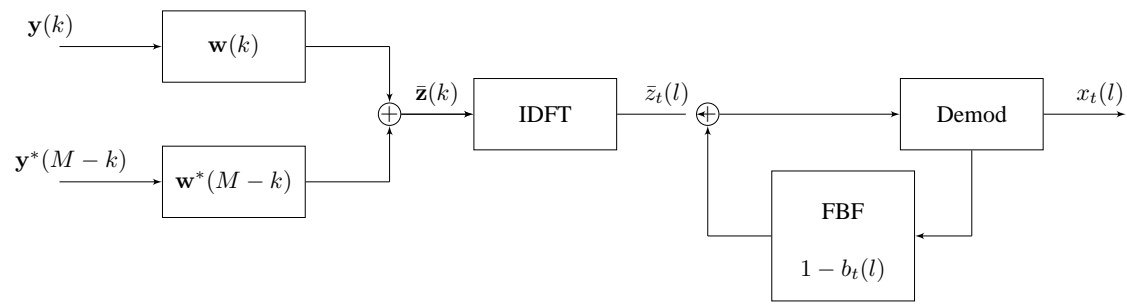

\begin{table}\label{Table-1}
\centering \caption{\texttt{Theoretically expected SNR gap of the receiver with respect to the  MFB in dB}}
\begin{tabular}{|c|c|c|}
\hline
\texttt{Receiver type} & \texttt{Gap for }$N_r=1$ & \texttt{Gap for} $N_r=2$\\
  \hline
 \texttt{ Conv ZF-LE} & \texttt{NA} & 3.0\\
 \hline
\texttt{Conv ZF-DFE} & 2.5 & 1.19\\
 \hline
\texttt{WL ZF-LE} & 3.0 & 1.25\\
  \hline
\texttt{WL ZF-DFE} & 1.17 & 0.5644\\
\hline
\end{tabular}
\end{table}

\begin{table}\label{Table-1}
\centering \caption{\texttt{SNR gap of the receiver with respect to MFB in dB at BER=0.01 for BPSK}}
\begin{tabular}{|c|c|c|}
\hline
\texttt{Receiver type} & \texttt{Gap for }$N_r=1$ & \texttt{Gap for} $N_r=2$\\
  \hline
 \texttt{ Conv ZF-LE} & \texttt{NA} & 3.0\\
 \hline
\texttt{Conv ZF-DFE} & 2.5 & 1.2\\
 \hline
 \texttt{Conv MMSE-LE} & 3.4 & 1.65 \\
 \hline
\texttt{Conv MMSE-DFE} & 1.4 & 0.75 \\
 \hline
\texttt{WL ZF-LE} & 3.2 & 1.3\\
  \hline
\texttt{WL ZF-DFE} & 1.2 & 0.6\\
\hline
\texttt{WL MMSE-LE} & 2.15 & 1.0 \\
  \hline
\texttt{WL MMSE-DFE} & 1.0 & 0.5 \\
\hline
\end{tabular}
\end{table}

\begin{table}\label{Table-1}
\centering \caption{\texttt{SNR gap of the receiver with respect to MFB in dB at BER=0.001 for BPSK}}
\begin{tabular}{|c|c|c|}
\hline
\texttt{Receiver type} & \texttt{Gap for } $N_r=1$ & \texttt{Gap for} $N_r=2$\\
  \hline
 \texttt{ Conv ZF-LE} & \texttt{NA} & 3.2\\
 \hline
\texttt{Conv ZF-DFE} & 2.6 &  1.1\\
\hline
\texttt{ Conv MMSE-LE} &  4.2 & 2.0 \\
 \hline
\texttt{Conv MMSE-DFE} &  1.7 & 0.8\\
 \hline
\texttt{WL ZF-LE} &  3.8 & 1.4\\
  \hline
\texttt{WL ZF-DFE} &  1.17 &  0.6\\
\hline
\texttt{WL MMSE-LE} &  2.55 &  1.0\\
  \hline
\texttt{WL MMSE-DFE} &  1.05 &  0.5\\
\hline
\end{tabular}
\end{table}

\begin{figure} [thb]
    \centerline{
        \epsfig{figure=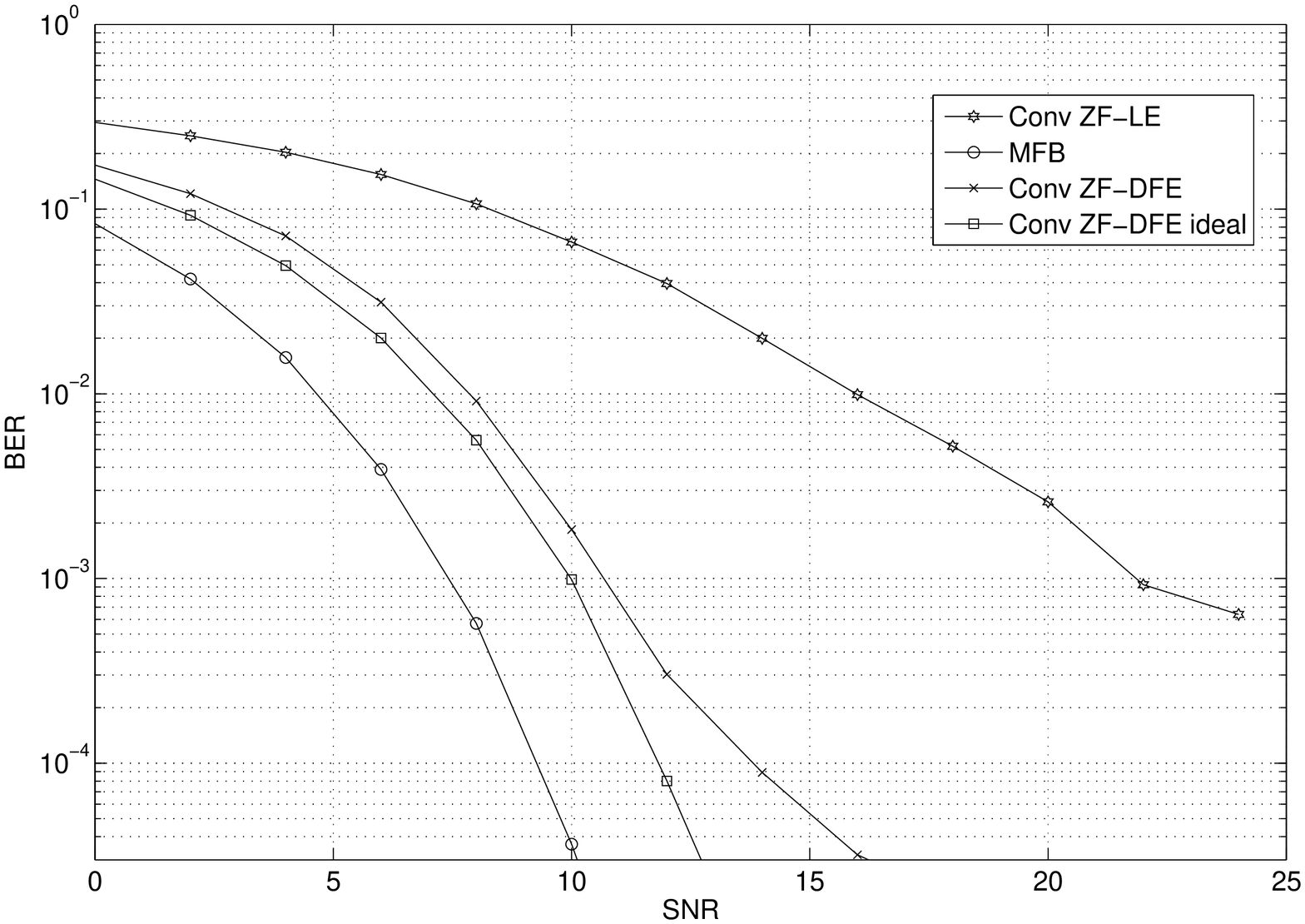, width=10.0cm, height=8.0cm}
    }
        \vspace*{0.0cm}  \caption{BPSK, Conventional ZF Receivers, L=20, $Nr=1$} \label{fig:FIGURE1}
\end{figure}

\begin{figure} [thb]
    \centerline{
        \epsfig{figure=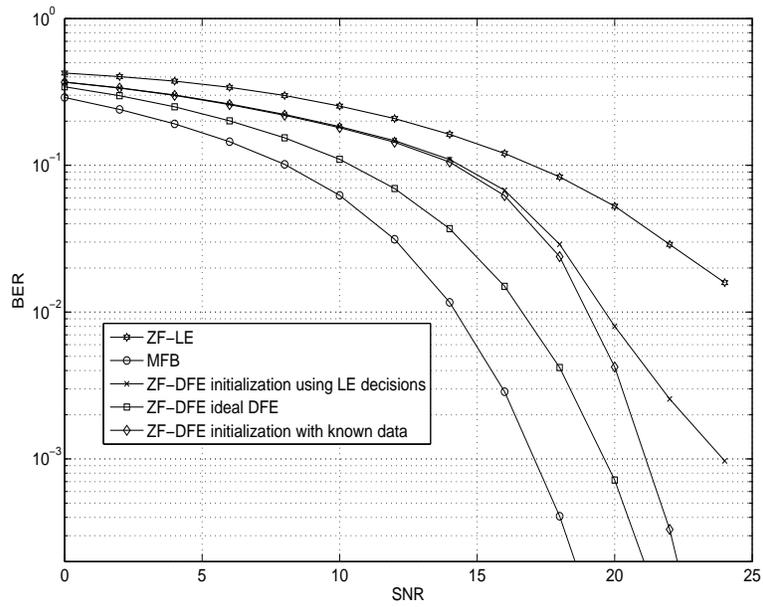, width=10.0cm, height=8.0cm}
    }
        \vspace*{0.0cm}  \caption{16-QAM, Conventional ZF Receivers, L=20, $Nr=1$} \label{fig:FIGURE2}
\end{figure}

\begin{figure} [thb]
    \centerline{
        \epsfig{figure=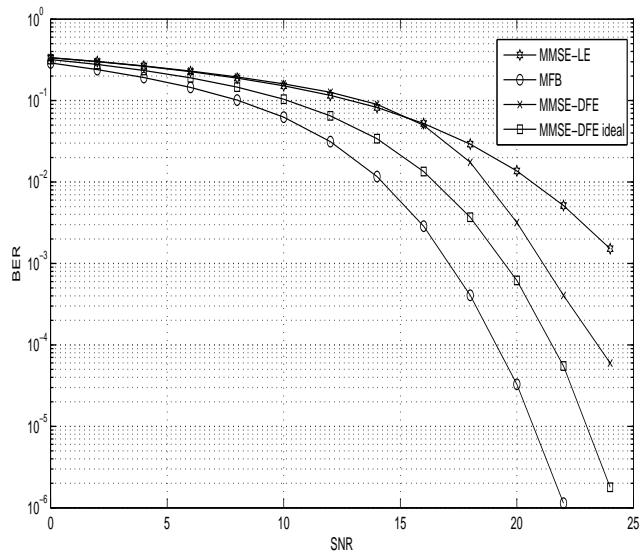, width=10.0cm, height=8.0cm}
    }
        \vspace*{0.0cm}  \caption{16-QAM, Conventional MMSE Receivers, L=20, $Nr=1$} \label{fig:FIGURE3}
\end{figure}

\begin{figure} [thb]
    \centerline{
        \epsfig{figure=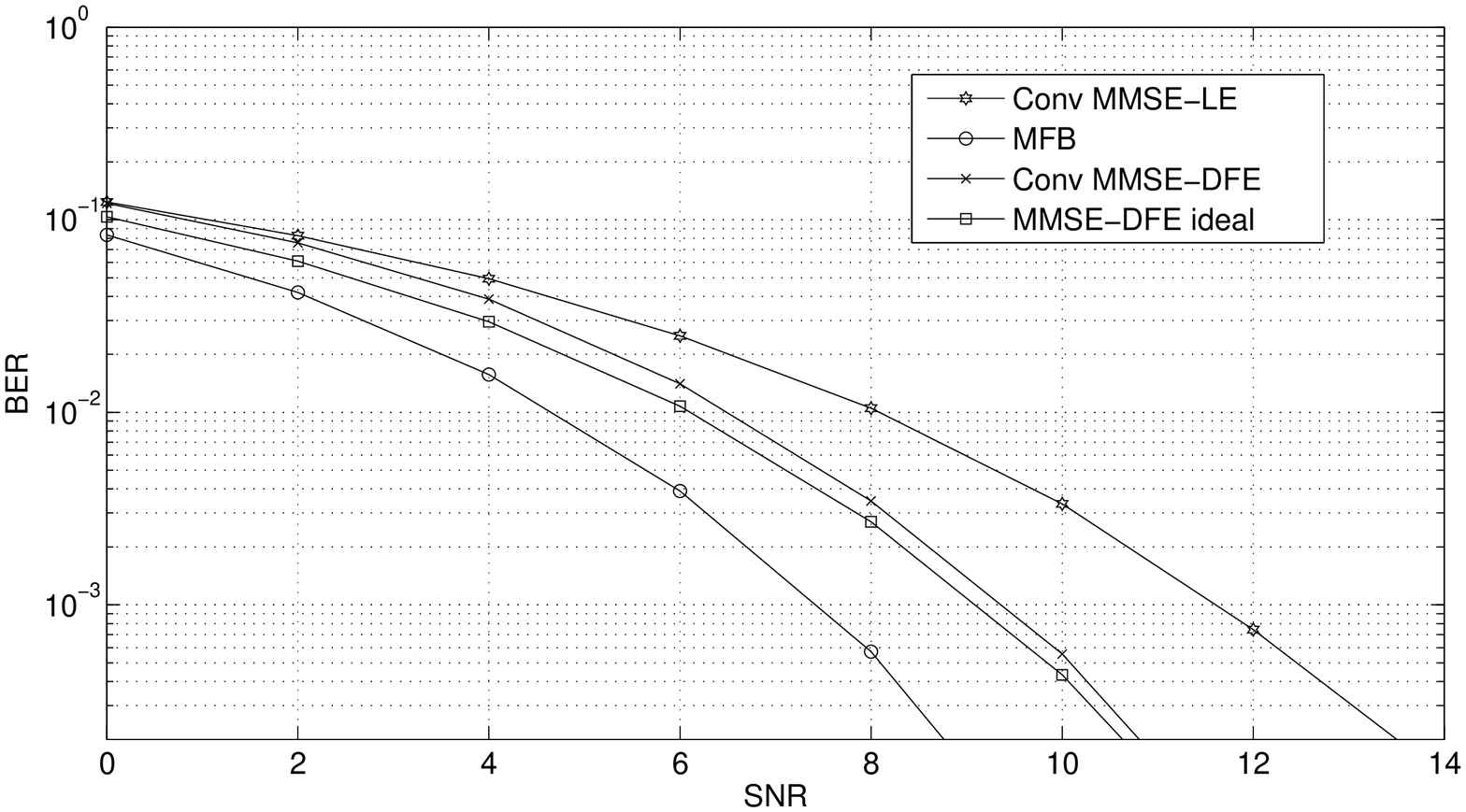, width=10.0cm, height=8.0cm}
    }
        \vspace*{0.0cm}  \caption{BPSK, Conventional MMSE Receivers, L=20, $Nr=1$} \label{fig:FIGURE4}
\end{figure}

\begin{figure} [thb]
    \centerline{
        \epsfig{figure=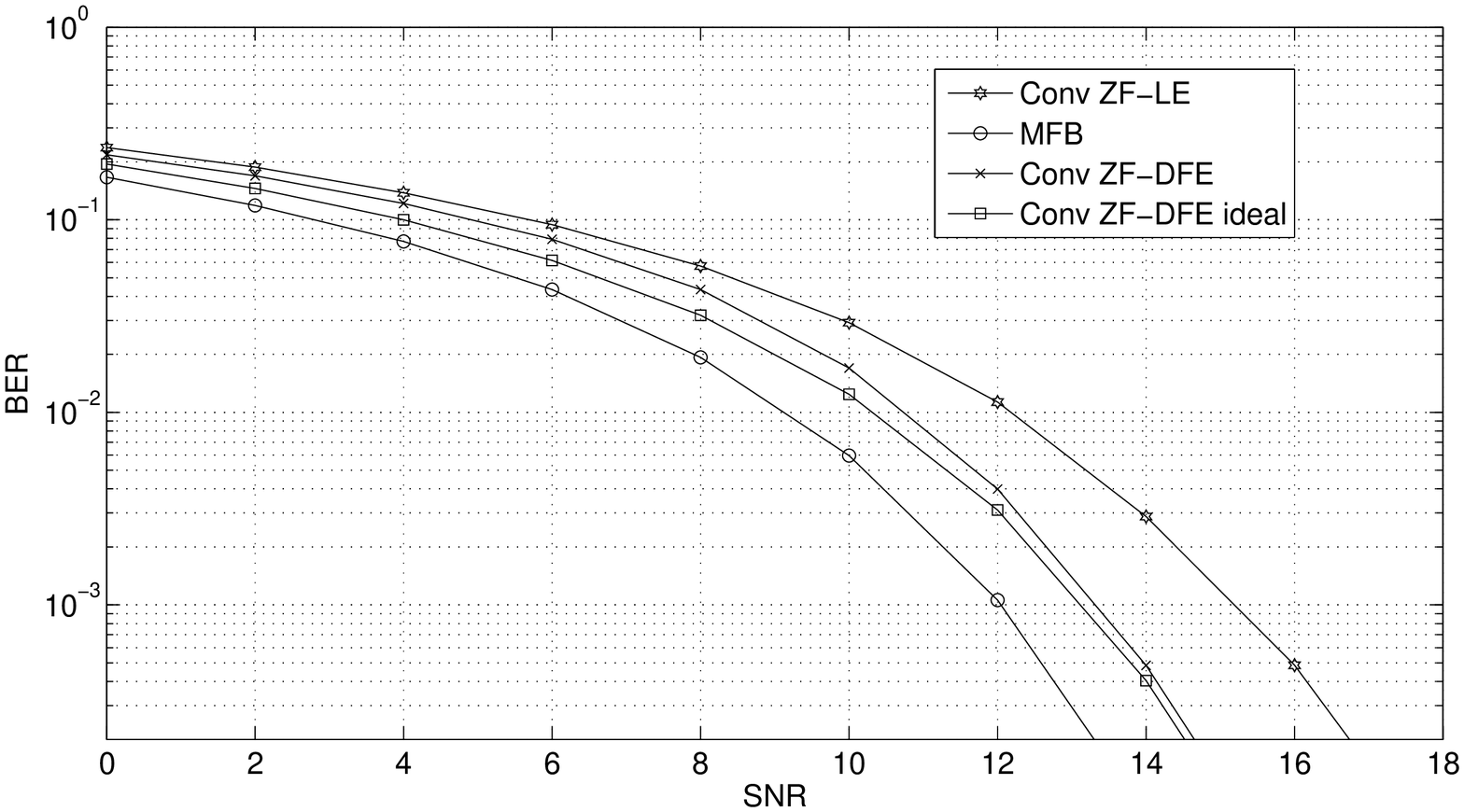, width=10.0cm, height=8.0cm}
    }
        \vspace*{0.0cm}  \caption{BPSK, Conventional ZF Receivers, L=20, $Nr=2$} \label{fig:FIGURE4a}
\end{figure}

\begin{figure} [thb]
    \centerline{
        \epsfig{figure=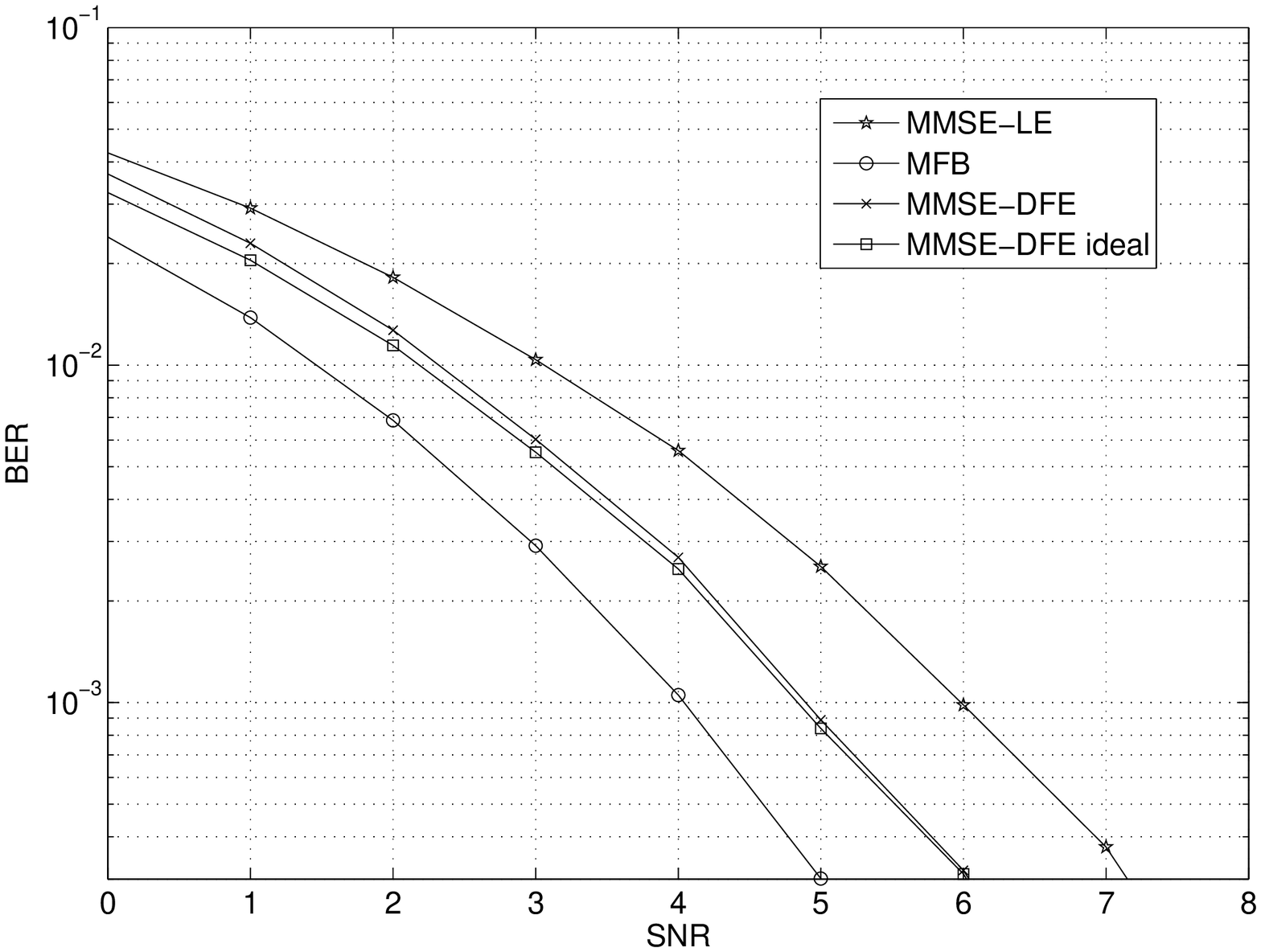, width=10.0cm, height=8.0cm}
    }
        \vspace*{0.0cm}  \caption{BPSK, Conventional MMSE Receivers, L=20, $Nr=2$} \label{fig:FIGURE4b}
\end{figure}

\begin{figure} [thb]
    \centerline{
        \epsfig{figure=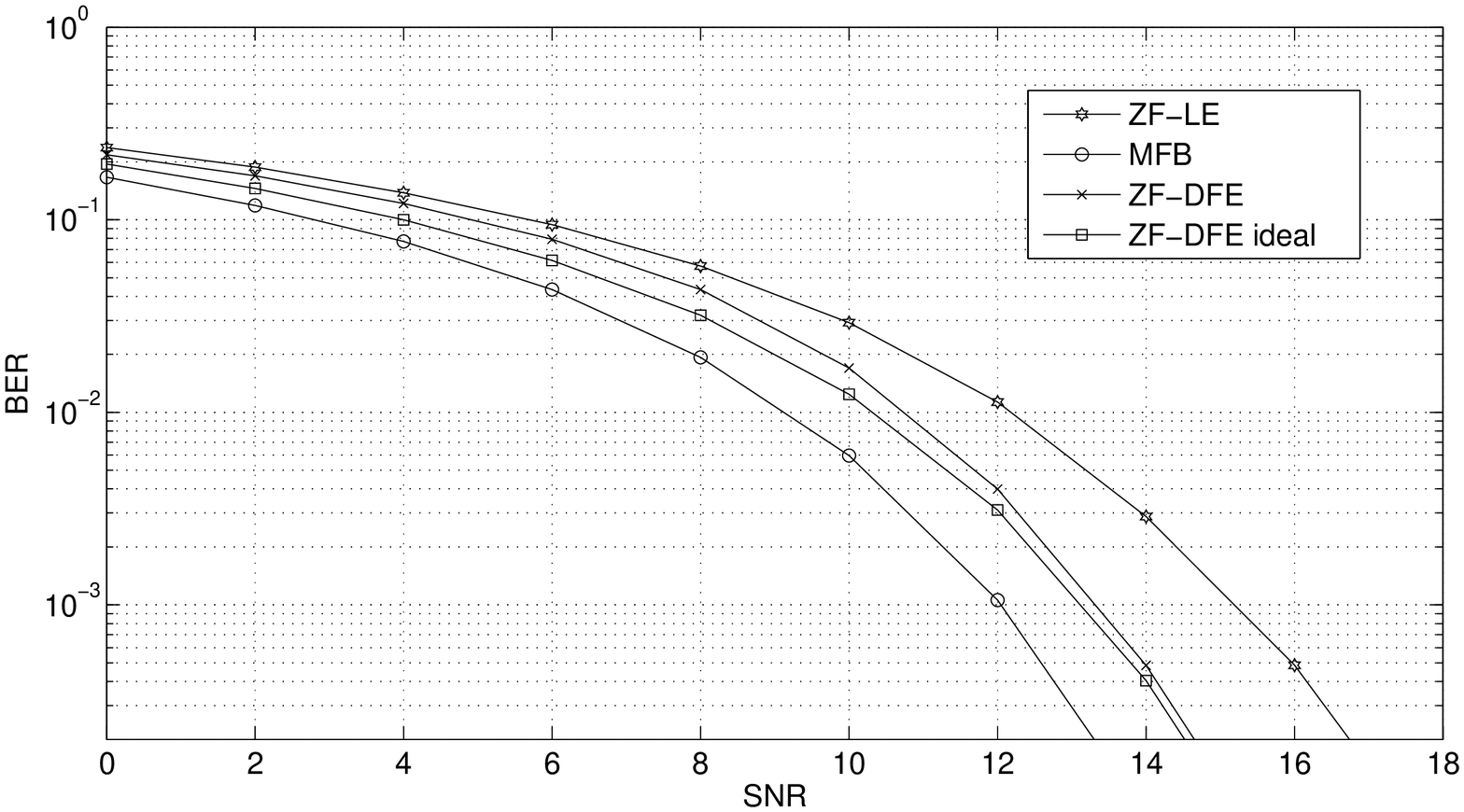, width=10.0cm, height=8.0cm}
    }
        \vspace*{0.0cm}  \caption{8-PSK, Conventional ZF Receivers, L=20, $Nr=2$} \label{fig:FIGURE6}
\end{figure}

\begin{figure} [thb]
    \centerline{
        \epsfig{figure=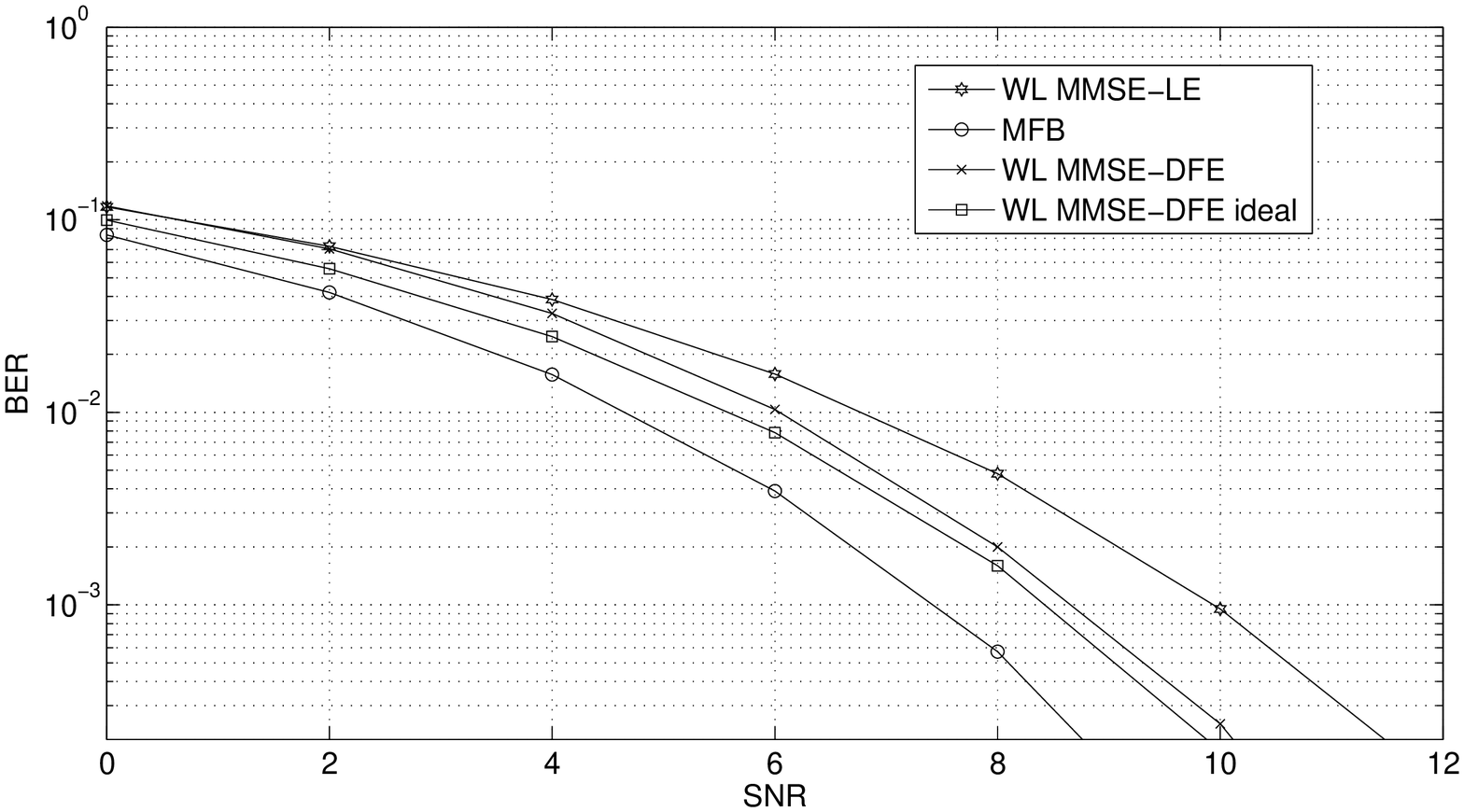, width=10.0cm, height=8.0cm}
    }
        \vspace*{0.0cm}  \caption{BPSK, WL MMSE Receivers, L=20, $Nr=1$} \label{fig:FIGURE5}
\end{figure}

\begin{figure} [thb]
    \centerline{
        \epsfig{figure=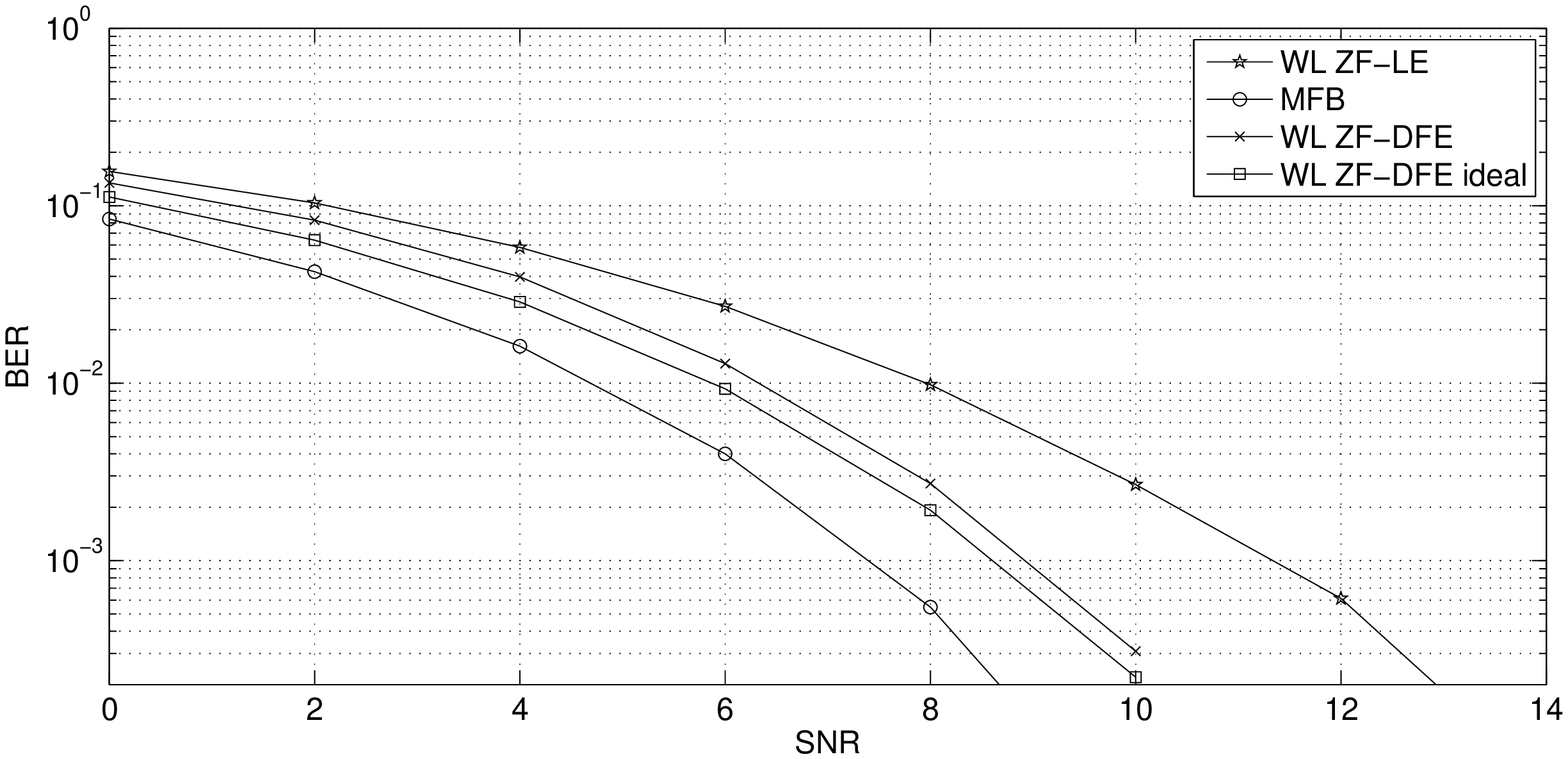, width=10.0cm, height=8.0cm}
    }
        \vspace*{0.0cm}  \caption{BPSK, WL ZF Receivers, L=20, $Nr=1$} \label{fig:FIGURE5a}
\end{figure}

\begin{figure} [thb]
    \centerline{
        \epsfig{figure=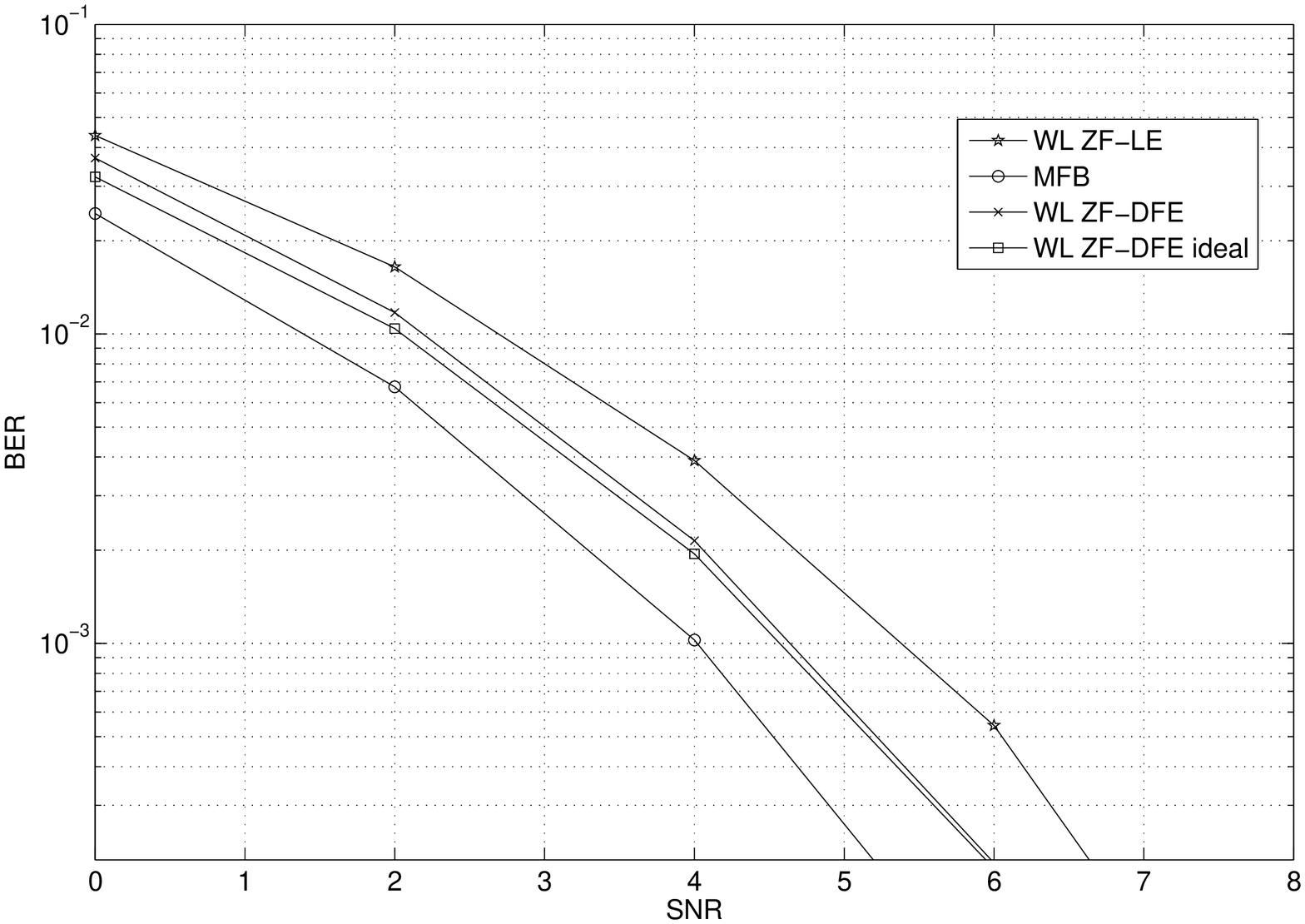, width=10.0cm, height=8.0cm}
    }
        \vspace*{0.0cm}  \caption{BPSK, WL ZF Receivers, L=20, $Nr=2$} \label{fig:FIGURE5b}
\end{figure}

\begin{figure} [thb]
    \centerline{
        \epsfig{figure=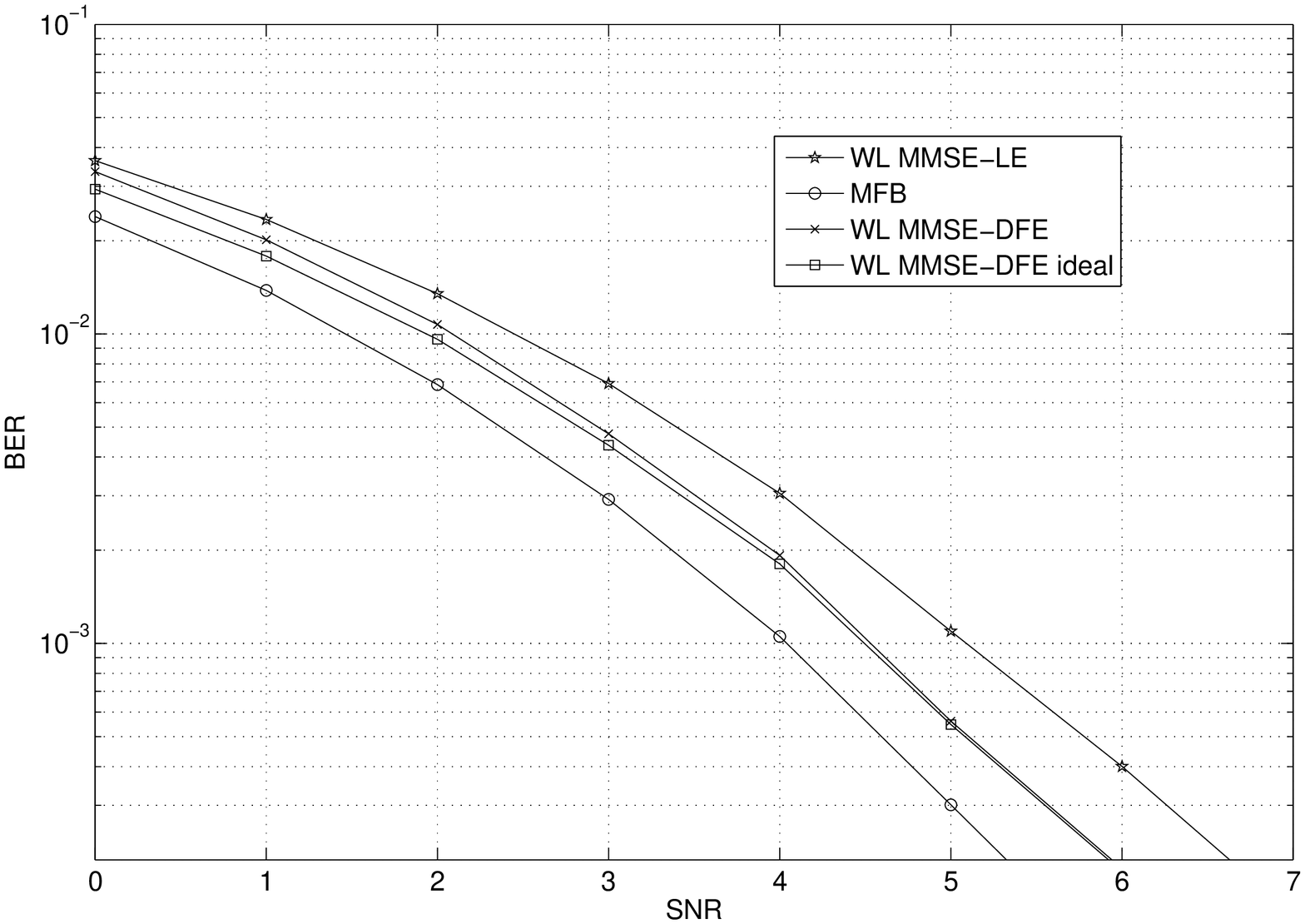, width=10.0cm, height=8.0cm}
    }
        \vspace*{0.0cm}  \caption{BPSK, WL MMSE Receivers, L=20, $Nr=2$} \label{fig:FIGURE7}
\end{figure}

\end{document}

%% file: formulaDef_08Aug2007.tex
\def\bGIQkm{\mathbf{\tilde{G}}^{m}_k}
\def\bC{\mathbf{C}}
\def\bV{\mathbf{V}}
\def\bh{\mathbf{h}}
\def\bw{\mathbf{w}}
\def\bW{\mathbf{W}}
\def\bo{\mathbf{1}}
\def\bs{\mathbf{s}}
\def\bv{\mathbf{v}}
\def\bg{\mathbf{g}}
\def\bG{\mathbf{G}}
\def\bs{\mathbf{s}}
\def\bc{\mathbf{c}}
\def\bn{\mathbf{n}}
\def\bU{\mathbf{U}}
\def\bl{\mathbf{l}}
\def\bk{\mathbf{k}}
\def\bthe{\mathbf{\theta}}
\def\bz{\mathbf{z}}
\def\by{\mathbf{y}}
\def\be{\mathbf{e}}
\def\bx{\mathbf{x}}
\def\bPk{\mathbf{P}_k}
\def\bPf{\mathbf{P} (f)}
\def\bQf{\mathbf{Q} (f)}
\def\bQof{\mathbf{Q_0} (f)}
\def\bDf{\mathbf{D} (f)}
\def\bDof{\mathbf{D_0} (f)}
\def\bDbf{\mathbf{\bar{D}} (f)}
\def\bPcf{\mathbf{P}^{\dagger} (f)}
\def\bh{\mathbf h}
\def\bp{\mathbf p}
\def\bH{\mathbf H}
\def\bP{\mathbf P}
\def\bX{\mathbf X}
\def\bY{\mathbf Y}
\def\bYt{\mathbf{Y}_t}
\def\bT{\mathbf T}
\def\bff{\mathbf f}
\def\bvareps{\bf \varepsilon}
\def\balpha{{\mathbf \alpha}}
\def\bA{\mathbf A}
\def\bbA{\mathbf A}
\def\bB{\mathbf B}
\def\bI{\mathbf I}
\def\bG{\mathbf G}
\def\bQ{\mathbf Q}
\def\bR{\mathbf R}
\def\byk{\mathbf{y}(k)}
\def\bhk{\mathbf{h}(k)}
\def\bgk{\mathbf{g}_{l}(k)}
\def\bw{\mathbf w}
\def\bwk{\mathbf{w}(k)}
\def\bzk{\mathbf{z}(k)}
\def\bnk{\mathbf{n}(k)}
\def\bxk{x(k)}
\def\bxhk{\hat{x}(k)}
\def\bek{\mathbf{e}(k)}
\def\be{\mathbf e}
\def\beIQ{\mathbf{\tilde{e}}}
\def\bbk{b(k)}
\def\bbbk{{\bar{b}}(k)}
\def\bGk{\mathbf{G}(k)}

\def\byIQk{\mathbf{\tilde{y}}(k)}
\def\byIk{\mathbf{\tilde{y}}_{I}(k)}
\def\byQk{\mathbf{\tilde{y}}_{Q}(k)}
\def\bzIQk{\mathbf{\tilde{z}}(k)}
\def\bzIQlek{\mathbf{\tilde{z}}_{\texttt{LE}}(k)}
\def\bhIQk{\mathbf{\tilde{h}}(k)}
\def\bgIQk{\mathbf{\tilde{g}}^{l}(k)}
\def\bwIQk{\mathbf{\tilde{w}}(k)}
\def\bwIk{\mathbf{\tilde{w}}_{I}(k)}
\def\bwQk{\mathbf{\tilde{w}}_{Q}(k)}
\def\bWIQk{\mathbf{\tilde{W}}(k)}
\def\bWIQdk{\mathbf{\tilde{W}}(k)}
\def\bWIQdl{\mathbf{\tilde{W}}(l)}
\def\bWIQlek{\mathbf{\tilde{W}_{\texttt{LE}}}(k)}
\def\bnIQk{\mathbf{\tilde{n}}(k)}
\def\bnIQck{\mathbf{\tilde{n}^{'}}({k+m})}
\def\bsIQk{\mathbf{\tilde{s}}(k)}
\def\bsIQck{\mathbf{\tilde{s}^{'}}({k+m})}
\def\bxIQk{\mathbf{\tilde{x}}(k)}
\def\bxIQhk{\mathbf{\hat{x}}(k)}
\def\beIQk{\mathbf{\tilde{e}}(k)}
\def\beIQck{\mathbf{\tilde{e}^{\dagger}}(k)}
\def\bbIQk{{\tilde{b}}(k)}
\def\bBIQk{\mathbf{\tilde{B}}(k)}
\def\bBIQbk{\mathbf{\bar{B}}(k)}
\def\beIQ{\mathbf{\tilde{e}}}
\def\bHIQ{\mathbf{\tilde{H}}}
\def\bhIQ{{\mathbf{ \tilde{h}}}}
\def\bHIQk{\mathbf{\tilde{H}}(k)}
\def\bhIQk{\mathbf{\tilde{h}}(k)}

\def\bhIQz{\mathbf{\tilde{h}} (z)}
\def\bhIQdz{\mathbf{\tilde{h}^{\dagger}} (z)}
\def\bhIQcz{\mathbf{\tilde{h}^{\dagger}} (z^{-1})}

\def\bhz{\mathbf{h} (z)}
\def\bhdz{\mathbf{h^{\dagger}} (z)}
\def\bhcz{\mathbf{h^{\dagger}} (z^{-1})}

\def\byIQ{\mathbf{\tilde{y}}}
\def\bwIQ{\mathbf{\tilde{w}}}
\def\bWIQ{\mathbf{\tilde{W}}}
\def\bnIQ{\mathbf{\tilde{n}}}
\def\bYIQf{\mathbf{\tilde{Y}} (f)}
\def\bYIQcf{\mathbf{\tilde{Y}^{\dagger}} (f)}
\def\byIQf{\mathbf{\tilde{y}} (f)}
\def\byIQcf{\mathbf{\tilde{y}^{\dagger}} (f)}
\def\bHIQf{\mathbf{\tilde{H}} (f)}
\def\bHIQcf{\mathbf{\tilde{H}^{\dagger}} (f)}
\def\bhIQf{\mathbf{\tilde{h}} (f)}
\def\bomegaIQf{\mathbf{\tilde{\omega}} (f)}
\def\bhIQdf{\mathbf{\tilde{h}^{\dagger}} (f)}
\def\bhIQcf{\mathbf{\tilde{h}^{\dagger}} (f)}
\def\bomegaIQcf{\mathbf{\tilde{\omega}^{\dagger}} (f)}
\def\bhIcf{\mathbf{h^{\dagger}}_{I} (f)}
\def\bhQcf{\mathbf{h^{\dagger}}_{Q} (f)}
\def\bgIQf{\mathbf{\tilde{g}}_{l}(f)}
\def\bgIQcf{{\mathbf{\tilde{g}}^{\dagger}}_l (f)}
\def\bGIQf{\mathbf{\tilde{G}}^{l}(f)}
\def\bGIQcf{{\mathbf{\tilde{G}}^{\dagger}}_l (f)}
\def\bWIQf{\mathbf{\tilde{W}} (f)}
\def\bwIQf{\mathbf{\tilde{w}} (f)}
\def\bwIf{\mathbf{w}_{I} (f)}
\def\bwQf{\mathbf{w}_{Q} (f)}
\def\bWIQdf{\mathbf{\tilde{W}_{\texttt{DFE}}} (f)}
\def\bWIQle{\mathbf{\tilde{W}_{\texttt{LE}}} (f)}
\def\bwIQle{\mathbf{\tilde{w}_{\texttt{LE}}} (f)}
\def\bNIQf{\mathbf{\tilde{n}} (f)}
\def\bNIQcf{\mathbf{\tilde{n}^{\dagger}} (f)}
\def\bnIQf{\mathbf{\tilde{n}} (f)}
\def\bXIQf{\mathbf{\tilde{X}} (f)}
\def\bEIQf{\mathbf{\tilde{E}} (f)}
\def\bEIQcf{\mathbf{\tilde{E}^{\dagger}} (f)}
\def\beIQf{\mathbf{\tilde{e}} (f)}
\def\beIQcf{\mathbf{\tilde{e}^{\dagger}} (f)}
\def\bEIQdf{\mathbf{\tilde{E}_{\texttt{dfe}}} (f)}
\def\bEIQ{\mathbf{\tilde{E}}}
\def\bBIQf{\mathbf{\tilde{B}}(f)}
\def\bBIQcf{\mathbf{\tilde{B}^{\dagger}}(f)}
\def\bYf{\mathbf {Y} (f)}
\def\byf{\mathbf {y} (f)}
\def\bycf{\mathbf {y^{\dagger}} (f)}
\def\bHf{\mathbf {H} (f)}
\def\bhf{\mathbf {h} (f)}
\def\bgf{\mathbf{g}^{m}(f)}
\def\bgcf{\mathbf{g}^{m\dagger} (f)}
\def\bhcf{\mathbf {h^{\dagger}} (f)}
\def\bWf{\mathbf {W} (f)}
\def\bwf{\mathbf {w} (f)}
\def\bWdf{\mathbf{W_{\texttt{DFE}}} (f)}
\def\bWle{\mathbf{W_{\texttt{LE}}} (f)}
\def\bwdf{\mathbf{w_{\texttt{DFE}}} (f)}
\def\bwle{\mathbf{w_{\texttt{LE}}}(f)}
\def\bbdf{{b_{\texttt{DFE}}}(f)}
\def\bnf{\mathbf{n} (f)}
\def\bEf{\mathbf{E} (f)}
\def\bEcf{\mathbf{E^{\dagger}} (f)}
\def\bxf{x(f)}
\def\bxhf{\hat{x}(f)}
\def\bef{\mathbf{e} (f)}
\def\becf{\mathbf{e^{\dagger}} (f)}
\def\bEdf{\mathbf{E_{\texttt{dfe}}} (f)}
\def\bedf{\mathbf{e_{\texttt{dfe}}} (f)}
\def\bE{\mathbfE}
\def\bBf{\mathbf{B} (f)}
\def\bBcf{\mathbf{B^{\dagger}} (f)}
\def\bbf{\mathbf{b} (f)}
\def\bbcf{\mathbf{b^{\dagger}} (f)}

\def\buf{u (f)}
\def\budf{{u^{\dagger}} (f)}
\def\buif{{u^{-1}} (f)}
\def\bucf{{u^{\dagger}} (f)}

\def\bYIQD{\mathbf{\tilde{Y}} (D)}
\def\bYIQcD{\mathbf{\tilde{Y}^{\dagger}} (D^{-*})}
\def\byIQD{\mathbf{\tilde{y}} (D)}
\def\byIQcD{\mathbf{\tilde{y}^{\dagger}} (D^{-*})}
\def\bGIQD{\mathbf{\tilde{G}^{m}} (D)}
\def\bHIQD{\mathbf{\tilde{H}} (D)}
\def\bHIQcD{\mathbf{\tilde{H}^{\dagger}} (D^{-*})}
\def\bhIQD{\mathbf{\tilde{h}} (D)}
\def\bhID{\mathbf{h}_I (D)}
\def\bhQD{\mathbf{h}_Q (D)}
\def\bhIQdD{\mathbf{\tilde{h}^{\dagger}} (D)}
\def\bhIQcD{\mathbf{\tilde{h}^{\dagger}} (D^{-*})}
\def\bhIcD{\mathbf{h^{\dagger}}_{I} (D^{-*})}
\def\bhQcD{\mathbf{h^{\dagger}}_{Q} (D^{-*})}
\def\bgIQD{\mathbf{\tilde{g}} (D)}
\def\bgIQcD{\mathbf{\tilde{g}^{\dagger}} (D^{-*})}
\def\bWIQD{\mathbf{\tilde{W}} (D)}
\def\bwIQD{\mathbf{\tilde{w}} (D)}
\def\bwID{\mathbf{w}_{I} (D)}
\def\bwQD{\mathbf{w}_{Q} (D)}
\def\bWIQdD{\mathbf{\tilde{W}} (D)}
\def\bWIQle{\mathbf{\tilde{W}_{\texttt{LE}}} (D)}
\def\bwIQle{\mathbf{\tilde{w}_{\texttt{LE}}} (D)}
\def\bnIQcD{\mathbf{\tilde{n}^{\dagger}} (D^{-*})}
\def\bnIQD{\mathbf{\tilde{n}} (D)}
\def\bxIQD{\mathbf{\tilde{x}} (D)}
\def\bxIQcD{\mathbf{\tilde{x}}^{\dagger} (D)}
\def\bEIQD{\mathbf{\tilde{E}} (D)}
\def\bEIQcD{\mathbf{\tilde{E}^{\dagger}} (D^{-*})}
\def\beIQD{\mathbf{\tilde{e}} (D)}
\def\beIQcD{\mathbf{\tilde{e}^{\dagger}} (D^{-*})}
\def\bEIQdD{\mathbf{\tilde{E}_{\texttt{dfe}}} (D)}
\def\bEIQ{\mathbf{\tilde{E}}}
\def\bBIQD{\mathbf{\tilde{B}}(D)}
\def\bBIQcD{\mathbf{\tilde{B}^{\dagger}}(D^{-*})}
\def\bYD{\mathbf {Y} (D)}
\def\byD{\mathbf {y} (D)}
\def\bycD{\mathbf {y^{\dagger}} (D^{-*})}
\def\bHD{\mathbf {H} (D)}
\def\bhD{\mathbf {h} (D)}
\def\bgD{\mathbf {g} (D)}
\def\bhcD{\mathbf {h^{\dagger}} (D^{-*})}
\def\bhIcD{\mathbf {h^{\dagger}}_{I} (D^{-*})}
\def\bhQcD{\mathbf {h^{\dagger}}_{Q} (D^{-*})}
\def\bgcD{\mathbf {g^{\dagger}} (D^{-*})}
\def\bWD{\mathbf {W} (D)}
\def\bwID{\mathbf {w}_{I} (D)}
\def\bwQD{\mathbf {w}_{Q} (D)}
\def\bwD{\mathbf {w} (D)}
\def\bWdD{\mathbf{W_{\texttt{DFE}}} (D)}
\def\bWle{\mathbf{W_{\texttt{LE}}} (D)}
\def\bwdD{\mathbf{w_{\texttt{DFE}}} (D)}
\def\bwle{\mathbf{w_{\texttt{LE}}}(D)}
\def\bbdD{{b_{\texttt{DFE}}}(D)}
\def\bnD{\mathbf{n} (D)}
\def\bncD{\mathbf {n^{\dagger}} (D^{-*})}
\def\bED{\mathbf{E} (D)}
\def\bEcD{\mathbf{E^{\dagger}} (D^{-*})}
\def\bxD{x(D)}
\def\bxhD{\hat{x}(D)}
\def\beD{\mathbf{e} (D)}
\def\becD{\mathbf{e^{\dagger}} (D^{-*})}
\def\bEdD{\mathbf{E_{\texttt{dfe}}} (D)}
\def\bedD{\mathbf{e_{\texttt{dfe}}} (D)}
\def\bE{\mathbfE}
\def\bBD{\mathbf{B} (D)}
\def\bBcD{\mathbf{B^{\dagger}} (D^{-*})}
\def\bbD{b (D)}
\def\bbcD{\mathbf{b^{\dagger}} (D^{-*})}
\def\bGD{\mathbf{G} (D)}
\def\bgD{\mathbf{g} (D)}

\def\buD{u (D)}
\def\budD{{u^{\dagger}} (D^{-*})}
\def\buiD{{u^{-1}} (D)}
\def\bucD{{u^{\dagger}} (D)}

\def\buIQD{{\tilde{u}} (D)}
\def\buIQcD{\tilde{u}^{\dagger} (D)}
\def\buIQiD{\tilde{u}^{-1} (D)}
\def\buIQdD{\tilde{u}^{\dagger} (D^{-*})}
\def\buIQdiD{{\tilde{u}^{\dagger{-1}}}(D^{-*})}

\def\bbIQD{\tilde{b} (D)}
\def\bbIQcD{\tilde{b}^{\dagger} (D^{-*})}
\def\bReeIQ{\mathbf{R_{\tilde{e}\tilde{e}}}}
\def\bRee{\mathbf{R_{ee}}}
\def\bReeIQi{\mathbf{R^{-1}_{\tilde{e}\tilde{e}}}}
\def\bRxxD{R_{xx} (D)}
\def\bRxixD{R_{x^{i}x^{i}} (D)}
\def\bRxixiD{R^{-1}_{x^{i}x^{i}} (D)}
\def\bRxyD{\mathbf{R_{xy}} (D)}
\def\bRyyD{\mathbf{R_{yy}} (D)}
\def\bRyyDi{\mathbf{R^{-1}_{yy}} (D)}
\def\bRnnD{\mathbf{R_{nn}} (D)}
\def\bRiiD{\mathbf{R_{ii}} (D)}
\def\bRinD{\mathbf{R_{(i+n)}} (D)}
\def\bRiniD{\mathbf{R^{-1}_{(i+n)}} (D)}
\def\bRnniD{\mathbf{R^{-1}_{nn}} (D)}
\def\bRxxiD{R^{-1}_{xx} (D)}
\def\bRyyiD{\mathbf{R^{-1}_{yy}} (D)}
\def\bRxxIQD{\mathbf{R_{\tilde{x}\tilde{x}}} (D)}
\def\bRxyIQD{\mathbf{R_{\tilde{x}\tilde{y}}} (D)}
\def\bRyyIQD{\mathbf{R_{\tilde{y}\tilde{y}}} (D)}
\def\bRnnIQD{\mathbf{R_{\tilde{n}\tilde{n}}} (D)}
\def\bRiiIQD{\mathbf{R_{\tilde{i}\tilde{i}}} (D)}
\def\bRiiIQf{\mathbf{R_{\tilde{i}\tilde{i}}} (f)}
\def\bRxxIQiD{\mathbf{R^{-1}_{\tilde{x}\tilde{x}}} (D)}
\def\bRnnIQiD{\mathbf{R^{-1}_{\tilde{n}\tilde{n}}} (D)}
\def\bRyyIQiD{\mathbf{R^{-1}_{\tilde{y}\tilde{y}}} (D)}

\def\bUIQD{\mathbf{\tilde{U}}(D)}
\def\bUIQiD{\mathbf{\tilde{U}^{-1}}(D)}
\def\bUIQcD{\mathbf{\tilde{U}^{\dagger}}(D)}
\def\bUIQdD{\mathbf{\tilde{U}^{\dagger}}(D^{-*})}
\def\bUIQdiD{\mathbf{\tilde{U}^{\dagger{-1}}}(D^{-*})}

\def\buID{\mathbf {u_{I}} (D)}
\def\buQD{\mathbf {u_{Q}} (D)}
\def\buIcD{\mathbf {u^{\dagger}_{I}} (D^{-*})}
\def\buQcD{\mathbf {u^{\dagger}_{Q}} (D^{-*})}
\def\buIQcD{\mathbf {\tilde{u}^{\dagger}} (D^{-*})}
\def\buIQdD{\tilde{u}^{\dagger} (D^{-*})}
\def\buIQdiD{\mathbf{\tilde{u}^{\dagger{-1}}}(D^{-*})}

\def\bUIQf{\mathbf{\tilde{U}}(f)}
\def\bUIQif{\mathbf{\tilde{U}^{-1}}(f)}
\def\bUIQcf{\mathbf{\tilde{U}^{\dagger}}(f)}
\def\bUIQdf{\mathbf{\tilde{U}^{\dagger}}(f)}
\def\bUIQdif{\mathbf{\tilde{U}^{\dagger{-1}}}(f)}

\def\buIQf{{\tilde{u}} (f)}
\def\buIQcf{\tilde{u}^{\dagger} (f)}
\def\buIQif{\tilde{u}^{-1} (f)}
\def\buIQdf{\tilde{u}^{\dagger} (f)}
\def\buIQdif{{\tilde{u}^{\dagger{-1}}}(f)}

\def\buIQz{{\tilde{u}} (z)}
\def\buIQcz{\tilde{u}^{\dagger} (z)}
\def\buIQiz{\tilde{u}^{-1} (z)}
\def\buIQdz{\tilde{u}^{\dagger} (z^{-1})}
\def\buIQdiz{{\tilde{u}^{\dagger{-1}}}(z^{-1})}

\def\bbIQf{\tilde{b} (f)}
\def\bbIQcf{\tilde{b}^{\dagger} (f)}
\def\bReeIQ{\mathbf{R_{\tilde{e}\tilde{e}}}}
\def\bRee{\mathbf{R_{ee}}}
\def\bReeIQi{\mathbf{R^{-1}_{\tilde{e}\tilde{e}}}}
\def\bReeIQl{\mathbf{\bar{R}_{\tilde{e}\tilde{e}}}}
\def\bReeIQli{\mathbf{\bar{R}^{-1}_{\tilde{e}\tilde{e}}}}
\def\bRxx{R_{xx} (f)}
\def\bRxix{R_{x^{i}x^{i}} (f)}
\def\bRxixi{R^{-1}_{x^{i}x^{i}} (f)}
\def\bRxy{\mathbf{R_{xy}} (f)}
\def\bRyy{\mathbf{R_{yy}} (f)}
\def\bRnn{\mathbf{R_{nn}}}
\def\bRii{\mathbf{R_{ii}} (f)}
\def\bRin{\mathbf{R_{(i+n)}} (f)}
\def\bRini{\mathbf{R^{-1}_{(i+n)}} (f)}
\def\bRnni{\mathbf{R^{-1}_{nn}} (f)}
\def\bRni{\mathbf{R^{-1}_{nn}}}
\def\bRnib{\mathbf{\bar{R}^{-1}_{nn}}}
\def\bRn{\mathbf{R_{nn}}}
\def\bRnb{\mathbf{\bar{R}_{nn}}}
\def\bRxxi{R^{-1}_{xx} (f)}
\def\bRyyi{\mathbf{R^{-1}_{yy}} (f)}
\def\bRxxIQ{\mathbf{R_{\tilde{x}\tilde{x}}} (f)}
\def\bRxyIQ{\mathbf{R_{\tilde{x}\tilde{y}}} (f)}
\def\bRyyIQ{\mathbf{R_{\tilde{y}\tilde{y}}} (f)}
\def\bRnnIQ{\mathbf{R_{\tilde{n}\tilde{n}}} (f)}
\def\bRnnIQb{\mathbf{\bar{R}_{\tilde{n}\tilde{n}}} (f)}
\def\bRiiIQ{\mathbf{R_{\tilde{i}\tilde{i}}} (f)}
\def\bRxxIQi{\mathbf{R^{-1}_{\tilde{x}\tilde{x}}} (f)}
\def\bRnnIQi{\mathbf{R^{-1}_{\tilde{n}\tilde{n}}} (f)}
\def\bRnnIQib{\mathbf{\bar{R}_{\tilde{n}\tilde{n}}} (f)}
\def\bRyyIQi{\mathbf{R^{-1}_{\tilde{y}\tilde{y}}} (f)}

\def\byIQbfk{\mathbf{\bar{y}} (f_k)}
\def\byIQfk{\mathbf{\tilde{y}} (f_k)}
\def\bhIQfk{\mathbf{\tilde{h}} (f_k)}
\def\bgIQfk{\mathbf{\tilde{g}} (f_k)}
\def\bnIQfk{\mathbf{\tilde{n}} (f_k)}
\def\bhIQcfk{\mathbf{\tilde{h}^{\dagger}} (f_k)}
\def\bgIQcfk{\mathbf{\tilde{g}^{\dagger}} (f_k)}
\def\bnIQcfk{\mathbf{\tilde{n}^{\dagger}} (f_k)}
\def\bRnnIQik{\mathbf{R^{-1}_{\tilde{n}\tilde{n}}} (f_k)}

\def\bUfm{\mathbf{U} (f_m)}
\def\bUcfm{\mathbf{U^{\dagger}} (f_m)}
\def\blamdafm{\mathbf{\Lambda} (f_m)}
\def\bhbIQfm{\mathbf{\bar{h}} (f_m)}
\def\bhbIQcfm{\mathbf{\bar{h}^{\dagger}} (f_m)}

\def\bhIQfm{\mathbf{\tilde{h}} (f_m)}
\def\bnIQfm{\mathbf{\tilde{n}} (f_m)}
\def\bhIQcfm{\mathbf{\tilde{h}^{\dagger}} (f_m)}
\def\bnIQcfm{\mathbf{\tilde{n}^{\dagger}} (f_m)}
\def\bRnnIQim{\mathbf{R^{-1}_{\tilde{n}\tilde{n}}} (f_m)}

\def\bh{{\mathbf h}}
\def\bE{{\mathbf E}}
\def\bg{{\mathbf g}}
\def\bomega{{\mathbf \omega}}
\def\bU{{\mathbf U}}
\def\bp{{\mathbf p}}
\def\bff{{\mathbf f}}
\def\bvareps{{\mathbf \varepsilon}}
\def\balpha{{\mathbf \alpha}}
\def\bA{{\mathbf A}}
\def\bB{{\mathbf B}}

\def\bh{{\mathbf h}}
\def\bp{{\mathbf p}}
\def\bH{{\mathbf H}}
\def\bP{{\mathbf P}}
\def\PIQ{{\tilde{ P}}}
\def\bff{{\mathbf f}}
\def\bvareps{\bf \varepsilon}
\def\balpha{{\mathbf \alpha}}
\def\bA{{\mathbf A}}
\def\bB{{\mathbf B}}
\def\bI{{\mathbf I}}
\def\bQ{{\mathbf Q}}

\def\bh{{\mathbf h}}
\def\bp{{\mathbf p}}
\def\bH{{\mathbf H}}
\def\bP{{\mathbf P}}
\def\bff{{\mathbf f}}
\def\bvareps{\bf \varepsilon}
\def\balpha{{\mathbf \alpha}}
\def\bA{{\mathbf A}}
\def\bB{{\mathbf B}}
\def\bI{{\mathbf I}}
\def\bQ{{\mathbf Q}}

\def\bSf{\mathbf {S} (f)}
\def\bLf{\mathbf {L} (f)}
\def\bLcf{\mathbf {L^{\dagger}} (f)}
\def\bD{\mathbf {D} }
\def\bSz{\mathbf {S} (z)}
\def\bLz{\mathbf {L} (z)}
\def\bLiz{\mathbf {L^{-1}} (z)}
\def\bLcz{\mathbf {L^{\dagger}} (z^{-1})}
\def\bL{\mathbf {L} }
\def\bZ{\mathbf {Z} }
\def\bP{\mathbf {P} }
\def\bE{\mathbf {E} }
\def\bKp{\mathbf {Kp} }
\def\bX{\mathbf {X} }
\def\bF{\mathbf {F} }
\def\bT{\mathbf {T} }
\def\bN{\mathbf {N} }
\def\bR{\mathbf {R} }
\def\bK{\mathbf {K} }
\def\bLi{\mathbf {Li} }
\def\bFt{\mathbf {Ft} }
